\documentclass[10pt,journal,letterpaper,twoside,twocolumn]{IEEEtran}
\usepackage{amsmath,epsfig}
\usepackage{setspace}
\usepackage{graphicx}
\usepackage{caption}
\usepackage{amsfonts,bm}
\usepackage{amssymb}
\usepackage{rotating}
\usepackage{color}
\usepackage{float}
\usepackage{multirow}
\usepackage[noadjust]{cite}

\graphicspath{{figures/}{.}}

\newcommand{\be}{\begin{equation}}
\newcommand{\ee}{\end{equation}}

\newcommand{\bb}{\boldsymbol}
\newcommand{\vx}{{\bb x}}
\newcommand{\vd}{{\bb d}}
\newcommand{\vh}{{\bb h}}
\newcommand{\mR}{{\bb R}}
\newcommand{\mI}{{\bb I}}
\newcommand{\mH}{{\bb H}}
\newcommand{\mS}{{\bb \Sigma}}
\newcommand{\vv}{{\bb v}}

\newcommand{\vz}{{\bb z}}

\newcommand{\vZ}{{\bb Z}}

\newcommand{\vo}{{\bb 0}}

\newcommand{\cH}[1]{\mathcal{H}_{#1}}

\newcommand{\ttop}{\!^\top\!}

\usepackage{color}  

\begin{document}

\singlespacing

\title{Technical Report: A New Decision-Theory-Based Framework for Echo Canceler Control}
\author{Tales Imbiriba,~\IEEEmembership{Member,~IEEE}, Jos\'{e}~Carlos~M.~Bermudez,~\IEEEmembership{Senior~Member,~IEEE}, \\Jean-Yves Tourneret,~\IEEEmembership{Senior~Member,~IEEE},
 and Neil J. Bershad,~\IEEEmembership{Fellow,~IEEE}
\thanks{Jean-Yves Tourneret is with the University of Toulouse, IRIT - ENSEEIHT - TéSA, 2 rue Camichel, BP 7122, 31071 Toulouse cedex 7, France, e-mail:Jean-Yves.Tourneret@enseeiht.fr. \newline Jos\'e Carlos Bermudez and Tales Imbiriba are with the Dept.\ of Electrical Engineering, Federal University of Santa Catarina, Florian\'opolis, SC, Brazil, e-mail:j.bermudez@ieee.org, and talesim@gmail.com. \newline Neil J. Bershad is with the Dept.\ of Electrical Engineering and Computer Science, University of California Irvine, Irvine, 1621 Santiago Drive, Newport Beach, CA 92660, U.S.A., e-mail:bershad@ece.uci.edu}
\bigskip \thanks{Corresponding author: J.-Y. Tourneret.}} 
\maketitle

\begin{abstract}
A control logic has a central role in many echo cancellation systems for optimizing the performance of adaptive filters while estimating the echo path. For reliable control, accurate double-talk (DT) and channel change (CC) detectors are usually incorporated to the echo canceler. This work expands the usual detection strategy to define a classification problem characterizing four possible states of the echo canceler operation. The new formulation allow the use of decision theory to continuously control the transitions among the different modes of operation. The classification rule reduces to a low cost statistics for which it is possible to determine the probability of error under all hypotheses, allowing the classification performance to be accessed analytically. Monte Carlo simulations using synthetic and real data illustrate the reliability of the proposed method. 
\end{abstract}

\begin{keywords}
Adaptive filters, adaptive signal processing, adaptive systems, echo cancellation, channel change, double-talk, classification, multivariate gamma distribution.
\end{keywords}




\section{Introduction} \label{sec:intro}

Echo cancellation is a requirement in modern voice communication systems. Speech echo cancelers (ECs) are employed in telephone networks (line echo cancelers) or in hands-free communications (acoustic echo cancelers). Most EC designs include two main blocks; a channel identification block  and a control logic block. The channel identification block tries to estimate the echo path, often employing adaptive filtering. However, the adaptive algorithm tends to diverge in the presence of near-end signals (double-talk -- DT). Hence, adaptation must be stopped during DT. On the other hand,  abrupt echo channel changes (CC) require a faster adaptation to improve tracking. Finally, in the absence of both DT and CC, a slow adaptation rate tends to improve channel estimation accuracy. The control logic is then required to control the transitions among these distinct modes of adaptive operation.  

Numerous approaches have been proposed to deal with DT or CC in echo cancelers. Some approaches have been proposed which do not require a DT detector, aiming at a continuous adaptation of the EC. These methods may employ signal correlations~\cite{benesty2000new}, independent component analysis (ICA)~\cite{Gunther2009,Gunther2015}, a continuous step size adjustment (variable step size)~\cite{Mader2000,GilCacho2014_2,Yang2017}, or a frequency domain approach~\cite{GilCacho2014}. However, eventual benefit of avoiding DT detection usually comes at the expense of decreased convergence rates~\cite{GilCacho2014_2}, and the need for additional information about loudspeaker-microphone coupling~\cite{Yang2017}, and about the near-end signal statistics~\cite{GilCacho2014}. Moreover, all such methods require extra estimation strategies resulting in extended memory usage and computational complexity, or significantly simplified implementations for practical applications.

Several works have proposed methods for DT detection in ECs without considerations regarding CC, such as~\cite{Ochiai1977,benesty2000new,schuldt2012delay,ikram2015double}. However, DT detection strategies that assume a static channel response may yield unpredictable performances in the presence of CC~\cite{ahgren2006study}. 

The vast majority of the techniques available for DT or CC detection rely on ad hoc statistics to make the decision, leading  to cumbersome design processes. A few works employ a statistical framework to formulate the detection problem. For instance, \cite{Carlemalm1996} proposes a maximum a posteriori (MAP) decision rule based on channel output observations and assuming Bernoulli distributed priors for the different hypotheses. A similar approach is used in \cite{Carlemalm1997}, but employing a Markov channel model. In \cite{Fozunbal2005}, a generalized likelihood ratio test (GLRT) is proposed using observations from both the channel input and output signals. DT and CC detection are considered. In \cite{jung2002new} and~\cite{jung2005}, a first test distinguishes single-talk from DT or CC, and a second test based on the echo path estimate detects DT. Though the latter two studies consider DT and CC in a single formulation, all these aforementioned statistical formulations have been proposed for the conventional adaptive EC structure \cite{Mader2000}.

\begin{figure}[bth]
\centering
\includegraphics[width=0.4\textwidth]{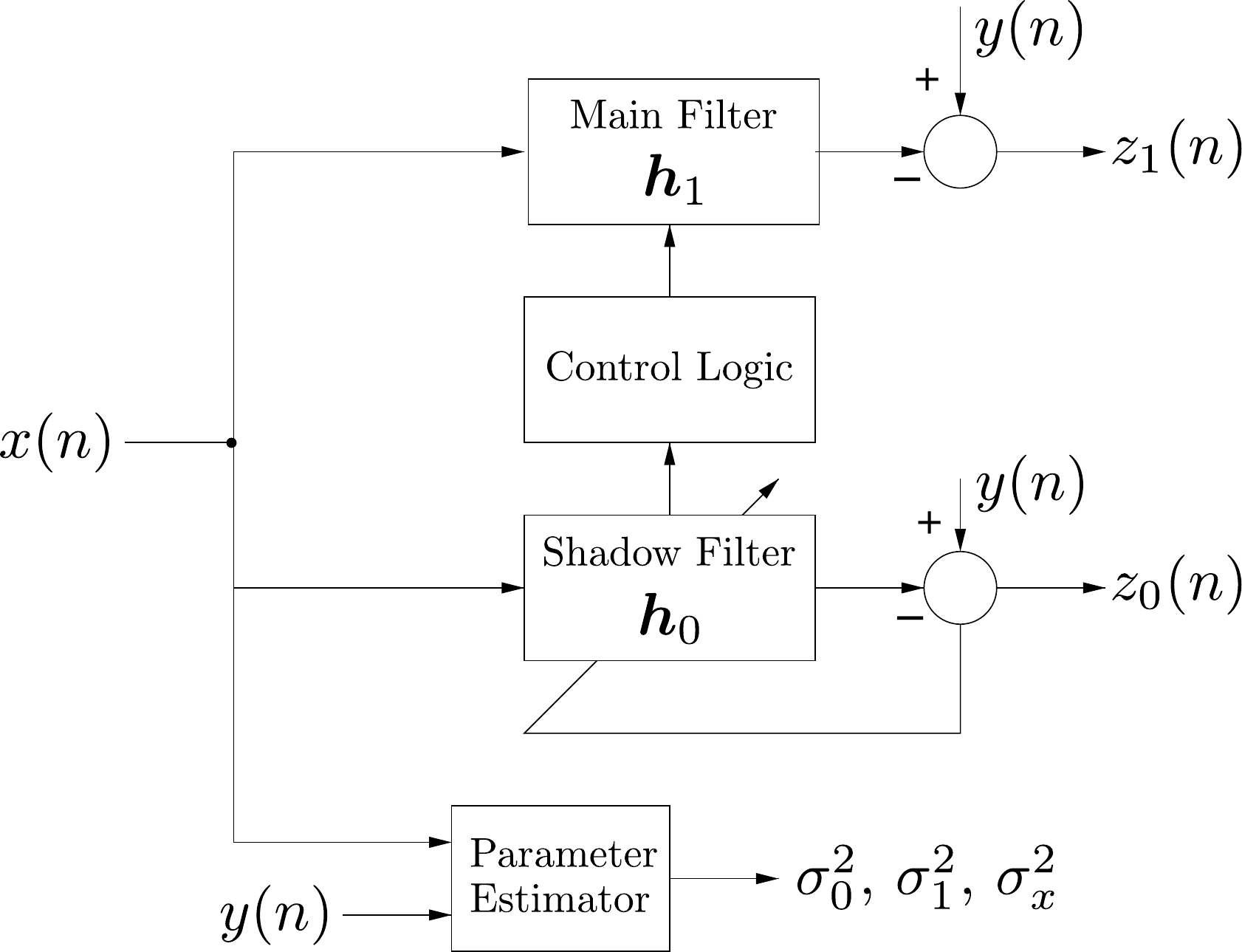} \caption{Basic echo
canceller structure.} \label{fig:Echo}
\end{figure}

An alternative EC structure has been proposed in \cite{Ochiai1977}, which uses a shadow adaptive filter that operates in parallel with the actual echo cancellation filter. The shadow filter coefficients are transferred to the echo cancellation filter when the shadow filter is a better estimate of the unknown channel response than the echo cancellation filter. From the authors experience, this structure allows a much better control of the EC convergence than the conventional structure. The EC structure is shown in Fig.~\ref{fig:Echo}. The EC consists of the main echo cancellation filter and the adaptive shadow filter. The output of the main filter is subtracted from the echo to obtain the canceled echo $z_1(n)$. The shadow filter weights are adapted continuously. The control logic is designed such that the shadow filter coefficients are copied to the main filter when this will improve the EC performance. A likelihood ratio test (LRT) detector based on the EC structure in Fig.~\ref{fig:Echo} was derived in~\cite{BershadTourneret2006} to detect DT versus CC. A generalized LRT (GLRT) that could be simplified to a sufficient statistic was proposed for the same EC structure in \cite{TourneretBershadBermudez2009}. The performance of the test statistic was evaluated as a function of the system parameters.  The idea developed in \cite{BershadTourneret2006} and \cite{TourneretBershadBermudez2009} was to use the detection result 1) to stop adaptation when DT was detected and 2) to adapt fast in the presence of channel change. The speed of adaptation was controlled by the adaptation step size. 

The decision theory-based DT and CC detection formulation in \cite{BershadTourneret2006,TourneretBershadBermudez2009} did not include decision theory based formulations for the exit from a DT or a CC condition. These decisions were still made in an ad hoc manner. 

This paper formulates the echo canceler control logic as a more general classification problem, with four hypotheses associated to the presence or absence of DT and to the presence or absence of CC
\begin{equation}
\begin{split}
&\mathcal{H}_0:
\text{no DT and no CC} \\
&\mathcal{H}_1: \text{no DT and CC} \\
&\mathcal{H}_2: \text{DT and no CC} \\
&\mathcal{H}_3: \text{DT and CC}.
\end{split}
\label{Neyman}
\end{equation}
There are several motivations for identifying these four classes. These motivations include 1) the possibility to adjust accurately the step size of the adaptive filter for long time intervals when there is no DT and no CC, resulting in smaller residual errors, 2) the inclusion of $\mathcal{H}_3$ adds an important degree of flexibility to the control logic that can be exploited, as will be shown in Section~\ref{sec:controlStrat}, 3) these four classes lead to a simple and low cost test statistic.

The paper is organized as follows. In Section \ref{sec:hypothesis} we introduce the signal models and derive the classification rules. In Section \ref{section3} we present the performance analysis of the proposed classifier. Monte Carlo simulations are presented in Section~\ref{sec:MC_Sim} to validate the theory. Section~\ref{sec:section5} discusses application of the proposed classification strategy and presents illustrative simulation results. Finally, Section \ref{section6} discusses the results and presents the conclusions.

\section{Double-talk and Channel Change Classification} \label{sec:hypothesis}

\subsection{Signal and Channel Models}
The channel input vector $\vx(n)=[x(n),\ldots,x(n-N+1)]^\top$ is of
dimension $N \times 1$ with covariance matrix $E [ \vx(n)
 \vx\ttop(n)]= \boldsymbol{\Sigma}_x$ and the channel output is a
scalar $y(n)$. The input signal is stationary within the decision
periods and the DT signal can be modelled by a white
Gaussian process for detection purposes \cite{Fozunbal2005}. Also,
$[y(n), \vx\ttop(n)]^\top$ is modelled as a zero-mean Gaussian vector. By
denoting as $\vh_0$ the adaptive shadow filter and $\vh_1$ as the
main echo cancellation filter, the channel output $y(n)$ can be
expressed as follows under the different hypotheses
\begin{equation}
\begin{split}
&\mathcal{H}_0 (\text{no DT, no CC}): y(n)=\vh_1^\top \vx(n)+n_0(n) \\
&\mathcal{H}_1 (\text{no DT, CC}): y(n)=\vh_0^\top \vx(n)+n_0(n) \\
&\mathcal{H}_2 (\text{DT, no CC}): y(n)=\vh_1^\top \vx(n)+n_0(n)+n_1(n)\\
&\mathcal{H}_3 (\text{DT, CC}): y(n)=\vh_0^\top \vx(n)+n_0(n)+n_1(n).
\end{split}
\label{Neyman2}
\end{equation}
The $\cH{0}$ hypothesis considers that $\vh_0$ has converged and has been recently copied to $\vh_1$. Hypothesis $\cH{1}$ assumes that $\vh_0$ has already converged after a channel change. In $\cH{2}$, a DT signal $n_1(n)$ happens after convergence of $\vh_0$ and copy to $\vh_1$ (similar to $\cH{0}$). Finally, a fourth hypothesis $\cH{3}$ considers that DT happens following a CC after $\vh_0$ has already converged to the new channel but has not yet been copied to $\vh_1$. All cases rely on the convergence (or divergence) of $\vh_0$ and its relation to $\vh_1$ resulting in several practical implications concerning the control logic block in Fig.~\ref{fig:Echo}. The control strategy will be addressed in Section~\ref{sec:section5}.

The additive noise
$n_0(n)$ is stationary zero-mean white\footnote{Note here that the
whiteness assumption for $n_0(n)$ is not restrictive since it is
always possible to whiten the channel outputs by pre-multiplying
consecutive samples by an appropriate matrix. Of course, this
operation assumes that the covariance matrix of consecutive noise
samples is known or can be estimated.} Gaussian, independent of
$\vx(n)$ with $E[n_0^2(n)]=\sigma_0^2$. The second additive noise
$n_1(n)$, modeling the DT, is zero-mean white Gaussian, and
independent of both $\vx(n)$ and $n_0(n)$ with
$E[n_1^2(n)]=\sigma_1^2$. Two error signals $z_0(n)=y(n)-\vh_0^\top
\vx(n)$ and $z_1(n)=y(n)-\vh_1^\top \vx(n)$ were introduced in
\cite{TourneretBershadBermudez2009} to facilitate the analysis.
These error signals can be expressed as follows under the different
hypotheses
 \begin{equation}
\begin{split}
&\!\!\!\mathcal{H}_0 (\text{no DT, no CC}):\\
&z_0(n)=(\vh_1-\vh_0)^\top \vx(n)+n_0(n),\, z_1(n)=n_0(n) \\
&\!\!\mathcal{H}_1 (\text{no DT, CC}):\\ 
&z_0(n)=n_0(n), \, z_1(n)=(\vh_0-\vh_1)^\top \vx(n)+n_0(n) \\
&\!\!\!\mathcal{H}_2 (\text{DT, no CC}):\\ 
&z_0(n)=(\vh_1-\vh_0)^\top \vx(n) +n_0(n)+n_1(n) \\ 
&z_1(n)=n_0(n)+n_1(n) \\
&\!\!\!\mathcal{H}_3 (\text{DT, CC}):\\ 
&z_0(n)=n_0(n)+n_1(n) \\ 
&z_1(n)=(\vh_0-\vh_1)^\top \vx(n) +n_0(n)+n_1(n).
\end{split}
\label{errors}
\end{equation}
\subsection{Classification Rule}
\label{subsec:ClassificationRule}
\subsubsection{One-Sample Case}
\begin{figure}[bth]
 \centering
 \includegraphics[width=0.25\textwidth]{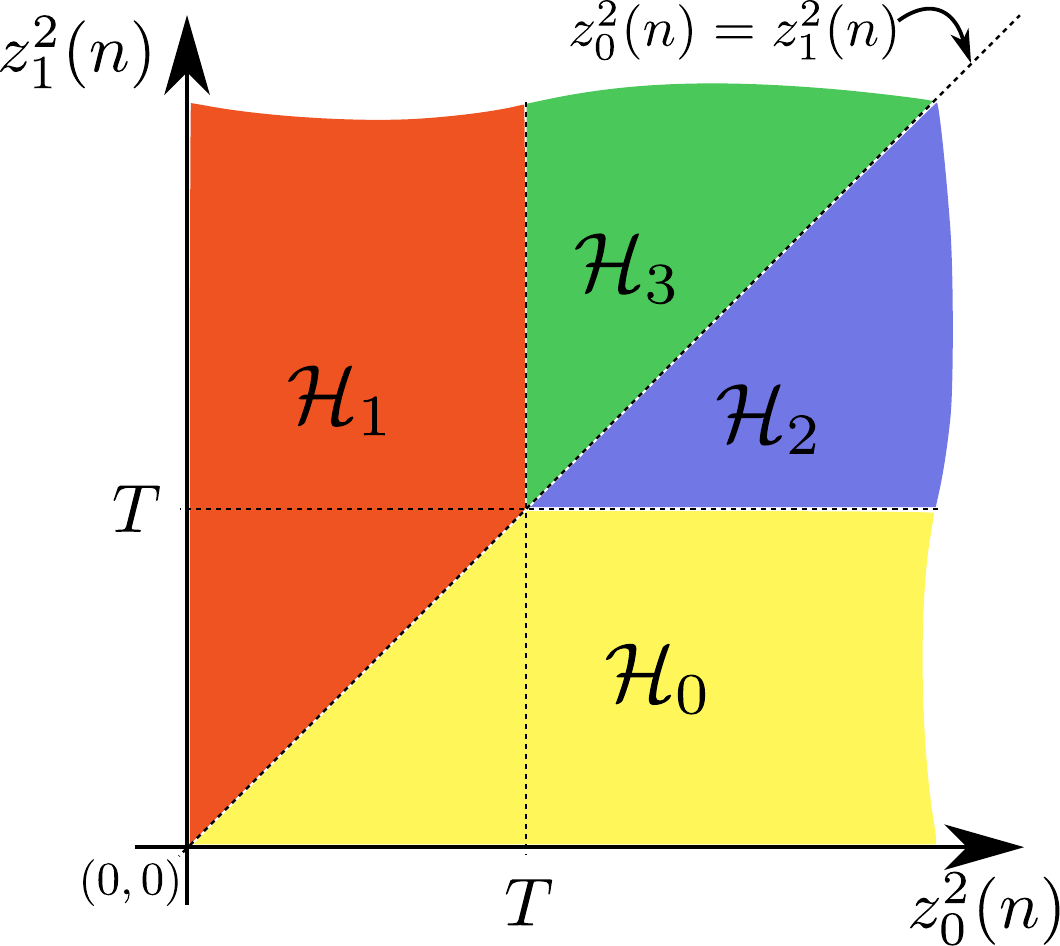}
 \caption{DT and CC Decision Regions in the $(z_0^2(n), z_1^2(n))$ Plane.}\label{fig:Decision}
\end{figure}

The joint pdf of $\vz(n)=[z_0(n),z_1(n)]^\top$ is Gaussian under all
hypotheses such that
\begin{equation}
p[\boldsymbol{z}(n)| \mathcal{H}_i] \sim
\mathcal{N}(\boldsymbol{0},\boldsymbol{\Sigma}_{i1}), \quad i=0,\dots,3
\label{lois}
\end{equation}
where the second subscript in $\boldsymbol{\Sigma}_{i1}$ ($1$ in
this case) indicates the $1$-sample case. The covariance matrices of
$\vz(n)$ under the different hypotheses can be written
\begin{equation} \label{Matrices01}
\begin{split}
&\boldsymbol{\Sigma}_{01}= \left(
\begin{array}{cc}
\sigma_0^2 + c_x^2& \sigma_0^2 \\
\sigma_0^2 & \sigma_0^2
\end{array}
\right)\\
&\boldsymbol{\Sigma}_{11}= \left(
\begin{array}{cc}
\sigma_0^2  & \sigma_0^2  \\
\sigma_0^2  & \sigma_0^2 + c_x^2
\end{array}
\right)
\end{split}
\end{equation}
\begin{equation} \label{Matrices11}
\begin{split}
 &\boldsymbol{\Sigma}_{21}= \left(
\begin{array}{cc}
\sigma_0^2 + \sigma_1^2+c_x^2 & \sigma_0^2 + \sigma_1^2 \\
\sigma_0^2 + \sigma_1^2 & \sigma_0^2 + \sigma_1^2
\end{array}
\right) \\
&\boldsymbol{\Sigma}_{31}= \left(
\begin{array}{cc}
 \sigma_0^2 + \sigma_1^2 & \sigma_0^2 + \sigma_1^2 \\
\sigma_0^2 + \sigma_1^2 & \sigma_0^2 + \sigma_1^2+c_x^2
\end{array}
\right)
\end{split}
\end{equation}
with
\begin{equation} \label{eq:cx}
c_x^2= (\vh_0 - \vh_1)^\top \mS_x (\vh_0 - \vh_1) 
\end{equation}
where $c_x^2$ can be interpreted as the power at the output of the difference filter with response $\vh_0-\vh_1$.
Assuming all hypotheses are equiprobable, the classification rule
minimizing the average probability of error decides hypothesis
$\mathcal{H}_i$ is true when
\begin{equation} \label{classif}
\begin{split}
&\frac{1}{\sqrt{|\mS_{i1}|}} \exp\left[ - \frac{1}{2} \vz\ttop(n)
\mS_{i1}^{-1} \vz(n) \right]\\ 
&\qquad\qquad\quad\;\geq \frac{1}{\sqrt{|\mS_{j1}|}}
\exp\left[ - \frac{1}{2} \vz\ttop(n) \mS_{j1}^{-1} \vz(n) \right]
\end{split}
\end{equation}
for all $j \neq i$. Equivalently, hypothesis $\mathcal{H}_i$ will be
accepted if
\begin{equation} \label{logclassif}
\boldsymbol{z}\ttop(n) \left( \boldsymbol{\Sigma}_{j1}^{-1} -
\boldsymbol{\Sigma}_{i1}^{-1} \right) \boldsymbol{z}(n)  \ge \ln
\left( \frac{|\mS_{i1}|}{|\mS_{j1}|}\right)
\end{equation}
for all $j \neq i$. Straightforward computations (detailed in Appendix~\ref{ClassificationRule}) allow one to compute the inverses and determinants of the
$2 \times 2$ matrices $\boldsymbol{\Sigma}_{i1}$ and
$\boldsymbol{\Sigma}_{j1}$ yielding the following classification
rule
\begin{equation}
\begin{split}
&\mathcal{H}_0 \; \textrm{aif} \; z_1^2(n) < z_0^2(n) \; \textrm{and} \; z_1^2(n) < T \\
&\mathcal{H}_1 \; \textrm{aif} \; z_1^2(n) > z_0^2(n) \; \textrm{and} \; z_0^2(n) < T \\
&\mathcal{H}_2 \; \textrm{aif} \; z_1^2(n) < z_0^2(n) \; \textrm{and} \; z_1^2(n) > T \\
&\mathcal{H}_3 \; \textrm{aif} \; z_1^2(n) > z_0^2(n) \; \textrm{and} \; z_0^2(n) > T
\end{split}
\label{simple}
\end{equation}
where ``aif'' means ``accepted if'' and 
\begin{equation}
T=\frac{\sigma_0^2 (\sigma_0^2+\sigma_1^2)}{\sigma_1^2} \ln \left( 1
+ \frac{\sigma_1^2}{\sigma_0^2} \right).
\end{equation}
The different decision regions corresponding to \eqref{simple} are illustrated in the $(z_0^2(n), z_1^2(n))$ plane in Fig.~\ref{fig:Decision}.

\subsubsection{Multiple Samples}\label{sec:MS}
The analysis above can be generalized to the case where multiple
time samples $\vz(n-k)$, for $k=p-1,\ldots,0$, are
available. The analysis is performed here for two samples (i.e.,
$p=2$) for simplicity and is generalized later. When two samples are
observed, the error signals $z_0(n),z_0(n-1)$ and $z_1(n),z_1(n-1)$
are considered. They can be expressed as follows under the different
hypotheses:

\noindent Under $H_0$:
\begin{equation} \label{msH0}
\begin{split}
&z_0(n)=(\vh_1-\vh_0)^\top \vx(n)+n_0(n)\\
&z_1(n)=n_0(n) \\
&z_0(n-1)=(\vh_1-\vh_0)^\top \vx(n-1)+n_0(n-1)\\
&z_1(n-1)=n_0(n-1).
\end{split}
\end{equation}
Under $H_1$:
\begin{equation} \label{msH1}
\begin{split}
&z_0(n)=n_0(n)\\
&z_1(n)=(\vh_0-\vh_1)^\top \vx(n)+n_0(n) \\
&z_0(n-1)=n_0(n-1)\\
&z_1(n-1)=(\vh_0-\vh_1)^\top \vx(n-1)+n_0(n-1).
\end{split}
\end{equation}
Under $H_2$:
\begin{equation} \label{msH2}
\begin{split}
&z_0(n)=(\vh_1-\vh_0)^\top \vx(n) +n_0(n)+n_1(n)\\
&z_1(n)=n_0(n)+n_1(n) \\
&z_0(n-1)=(\vh_1-\vh_0)^\top \vx(n-1)\\
&\qquad\qquad\quad  +n_0(n-1)+n_1(n-1))\\
&z_1(n-1)=n_0(n-1)+n_1(n-1).
\end{split}
\end{equation}
Under $H_3$:
\begin{equation} \label{msH3}
\begin{split}
&z_0(n)=n_0(n)+n_1(n)\\
&z_1(n)=(\vh_0-\vh_1)^\top \vx(n) +n_0(n)+n_1(n) \\
&z_0(n-1)=n_0(n-1)+n_1(n-1))\\
&z_1(n-1)=(\vh_0-\vh_1)^\top \vx(n-1)\\
&\qquad\qquad\quad  +n_0(n-1)+n_1(n-1).
\end{split}
\end{equation}
Defining $\vz_{2d}(n)=[z_0(n),z_0(n-1),z_1(n),z_1(n-1)]^\top$, $\vz_{2d}(n)$ is a
zero-mean Gaussian vector under all hypotheses. Straightforward
computations yield the covariance matrices of
$\vz_{2d}(n)$ under the different hypotheses. These matrices
can be expressed as
\begin{equation} \label{eq:M1}
\begin{split}
&\boldsymbol{\Sigma}_{02}= \left(
\begin{array}{cc}
\sigma_0^2 \mI_2 + \mH_x& \sigma_0^2 \mI_2 \\
\sigma_0^2 \mI_2 & \sigma_0^2 \mI_2
\end{array}
\right) \\
&\boldsymbol{\Sigma}_{12}= \left(
\begin{array}{cc}
\sigma_0^2 \mI_2  & \sigma_0^2 \mI_2 \\
\sigma_0^2 \mI_2  & \sigma_0^2 \mI_2 + \mH_x
\end{array}
\right)
\end{split}
\end{equation}
\begin{equation} \label{eq:M2}
\begin{split}
&\boldsymbol{\Sigma}_{22}= \left(
\begin{array}{cc}
(\sigma_0^2  + \sigma_1^2)\mI_2 +\mH_x & (\sigma_0^2  + \sigma_1^2)\mI_2 \\
(\sigma_0^2  + \sigma_1^2)\mI_2 & (\sigma_0^2  + \sigma_1^2)\mI_2
\end{array}
\right) \\
&\boldsymbol{\Sigma}_{32}= \left(
\begin{array}{cc}
 (\sigma_0^2  + \sigma_1^2)\mI_2 & (\sigma_0^2  + \sigma_1^2)\mI_2 \\
(\sigma_0^2  + \sigma_1^2)\mI_2 & (\sigma_0^2  +
\sigma_1^2)\mI_2+\mH_x
\end{array}
\right)
\end{split}
\end{equation}
where $\mI_2$ is the $2\times 2$ identity matrix and $\mH_x$ is given by Equation~\eqref{eq:Hx}.
\begin{figure*}[bth]
\begin{equation} \label{eq:Hx}
\mH_x= \left(
\begin{array}{cc}
\vh_0-\vh_1 & \vo \\ \vo & \vh_0-\vh_1 \end{array} \right)^{\!\top}\! \left(
\begin{array}{cc}
 \mS_x & \mR_{1x}  \\
 \mR_{-1x}  & \mS_x \end{array}
\right)
\left(
\begin{array}{cc}
\vh_0-\vh_1 & \vo \\
\vo & \vh_0-\vh_1 \end{array} \right)
\end{equation}
\end{figure*}
In~\eqref{eq:Hx}, $\mS_x = E[\vx(n)\vx\ttop(n)]$, $\mR_{1x}=E[\vx(n) \vx\ttop(n-1)]$, and $\mR_{-1x}=E[\vx(n-1) \vx\ttop(n)]$. The determinants and inverses of these block matrices
can be computed following \cite[p. 572]{Kay1993}
\begin{equation}
\begin{split}
&|\mS_{02}|=|\mS_{12}|=\sigma_0^4 |\mH_x|\\
&|\mS_{22}|=|\mS_{32}|=(\sigma_0^2 + \sigma_1^2)^2 |\mH_x|
\end{split}
\end{equation}
and
\begin{equation} \label{invM0}
\begin{split}
&\mS_{02}^{-1}= \left(
\begin{array}{cc}
 \mH_x^{-1} & -\mH_x^{-1} \\
-\mH_x^{-1} &  \frac{1}{\sigma_0^2} \mI_2 +\mH_x^{-1}
\end{array}
\right) \\
&\mS_{12}^{-1}= \left(
\begin{array}{cc}
\frac{1}{\sigma_0^2} \mI_2 +\mH_x^{-1} & -\mH_x^{-1} \\
-\mH_x^{-1} &  \mH_x^{-1}
\end{array}
\right)
\end{split}
\end{equation}
\begin{equation} \label{invM1}
\begin{split}
&\mS_{22}^{-1}= \left(
\begin{array}{cc}
 \mH_x^{-1} & -\mH_x^{-1} \\
-\mH_x^{-1} &  \frac{1}{\sigma_0^2+\sigma_1^2} \mI_2 +\mH_x^{-1}
\end{array}
\right) \\
&\mS_{32}^{-1}= \left(
\begin{array}{cc}
\frac{1}{\sigma_0^2+\sigma_1^2} \mI_2 +\mH_x^{-1} & -\mH_x^{-1} \\
-\mH_x^{-1} &  \mH_x^{-1}
\end{array}
\right).
\end{split}
\end{equation}
where $\mH_x^{-1}$ is assumed to exist.

Performing the same computations shown in Appendix~\ref{ClassificationRule} for vector $\vz_{2d}(n)$ and matrices \eqref{eq:M1} and \eqref{eq:M2}, the following multiple sample classification rule can then be obtained
\begin{equation}
\begin{split}
&\mathcal{H}_0\;\text{aif} \; \| \vz_1(n) \|^2 < \| \vz_0(n) \|^2 \;\text{and}\; \| \vz_1(n) \|^2 < T_2, \\
&\mathcal{H}_1\;\text{aif} \; \| \vz_1(n) \|^2 > \| \vz_0(n) \|^2 \;\text{and}\; \| \vz_0(n) \|^2 < T_2, \\
&\mathcal{H}_2\;\text{aif} \; \| \vz_1(n) \|^2 < \| \vz_0(n) \|^2 \;\text{and}\; \| \vz_1(n) \|^2 > T_2, \\
&\mathcal{H}_3\;\text{aif} \; \| \vz_1(n) \|^2 > \| \vz_0(n) \|^2 \;\text{and}\; \| \vz_0(n) \|^2 > T_2,
\end{split}
\label{eq:Decms}
\end{equation}
where $\vz_i(n)=[z_i(n), z_i(n-1)]^\top$, $\| \vz_i(n) \|^2=z_i^2(n)+z_i^2(n-1)$ and
\begin{equation}
T_2 = 2T= 2\frac{\sigma_0^2 (\sigma_0^2+\sigma_1^2)}{\sigma_1^2} \ln \left( 1
+ \frac{\sigma_1^2}{\sigma_0^2} \right).
\label{eq:T2}
\end{equation}

The factor $2$ multiplying $T$ in~\eqref{eq:T2} results from  $\ln
\left(|\mS_{i1}|/|\mS_{j1}|\right) = -2 \ln \left(1+ \sigma_1^2/\sigma_0^2 \right)$.
This result can be compared with \eqref{simple} obtained for the
one-sample case. The generalization to more than two samples is
straightforward. Indeed, in the $p$-sample case, the covariance
matrices $\mS_{ip}$ of $\vz_{pd}(n)$ are defined as in \eqref{eq:M1} and
\eqref{eq:M2}, with $\mI_2$ replaced with $\mI_p$,
and $\mH_x$ defined differently. However, since $\mH_x$ cancels
from the difference between the two inverses, the classification
rule for the $p$-sample case is expressed by \eqref{eq:Decms} with $\|
\vz_i(n) \|^2=\vz_i\ttop(n)\vz_i(n) =\sum_{k=0}^{p-1} z_i^2(n-k)$
the squared norm of $\vz_i(n)$, $i=0,1$, and with $T_2 = 2T$ replaced with $T_p = pT$.

\section{Performance Analysis} \label{section3}
This section studies the probability of classification error for the classifier proposed in Section~\ref{sec:hypothesis}.
\subsection{One-Sample Case}
It is clear from the classification rules \eqref{simple} that $\vd(n) = [z_0^2(n),z_1^2(n)]^\top$ is a sufficient statistic for the classification problem. Interestingly, the exact distribution of
$\vd(n)$ can be derived under all hypotheses, allowing for an analytical study of the classifier performance. First, we note that the elements of $\vd(n)$ form the diagonal of the matrix $\vZ = \vz(n) \vz\ttop(n)$. Now, since $\vz(n) = [z_0(n), z_1(n)]^\top$ is jointly distributed according to a zero-mean Gaussian distribution with covariance matrix $\boldsymbol{\Sigma}_{i1}$, see~\eqref{lois}, it is shown in Appendix~\ref{MultivariateGamma} that, under all hypotheses $\mathcal{H}_i$, $i=0,\dots,3$, $\vd(n)$ is distributed according to a multivariate gamma distribution denoted $\mathcal{G}(q,P)$ with shape parameter $q=p/2$ and scale parameter $P = \{p_1,p_2,p_{12}\}$, with 
%
\begin{equation}
\begin{split}
&p_1 = 2\bb{\Sigma}_{i,1}(1,1)\\
&p_2 = 2\bb{\Sigma}_{i,1}(2,2)\\
&p_{12} = 4\left[\bb{\Sigma}_{i,1}(1,1)\bb{\Sigma}_{i,1}(2,2) -\bb{\Sigma}_{i,1}(1,2)\bb{\Sigma}_{i,1}(2,1)\right]
\end{split}
\label{eq:Ptext}
\end{equation}
where $\bb{\Sigma}_{i,1}(1,1)$, $\bb{\Sigma}_{i,1}(1,2) = \bb{\Sigma}_{i,1}(2,1)$ and $\bb{\Sigma}_{i,1}(2,2)$ are the elements of the covariance matrix $\bb{\Sigma}_{i,1}$.

\subsection{Multiple-Sample}\label{sec:ProbMulti-Samp}

Once again it is clear that the vector $\vd(n) = [\|\vz_0(n)\|^2 ,  \|\vz_1(n)\|^2 ]^\top$ is a sufficient statistic for solving the proposed classification problem. 
Noting that $\vz_{pd}(n)$ is a rearrangement of the $p$ vectors $\vz(n-k)$, $k=0,\ldots,p-1$, the distribution of $\vd(n)$ can be obtained following the reasoning presented in Appendix~\ref{MultivariateGamma}, under the assumption of independence of vectors $\vz(n-i)$ and $\vz(n-j)$, $i\neq j$, and stationarity for $\vz(n-k)$.
Assuming the vectors $\vz(n-k)$, $k=0,\ldots,p-1$, to be distributed according to the same zero-mean Gaussian distribution with covariance matrix $\bb{\Sigma}_{i1}$, matrix $\bb{A} =\sum_{k=0}^{p-1}\vz(n-k)\vz\ttop(n-k)$ is distributed according to a Wishart distribution $\mathcal{W}_2(p,\bb{\Sigma}_{i1})$ with $p$ degrees of freedom~\cite[Theorem 3.2.4, pg. 91]{muirhead2005}.
Thus, $\vd(n) = \text{diag}(\bb{A})$ is distributed according to a multivariate gamma distribution with shape parameter $q=p/2$ and $P$ given by~\eqref{eq:P}. 


\subsection{Probability of Error}
To simplify the notation, define $t_0$ and $t_1$ such that $\vd(n) = [\|\vz_0(n)\|^2 ,  \|\vz_1(n)\|^2 ]^\top = [t_0,t_1]^\top$. Also consider $f_j$ to be the bivariate gamma density associated with hypothesis $\cH{j}$.
Then, the probability of error $P_{ij}=P(\mathcal{H}_i|\mathcal{H}_j)$, can be computed as:
\begin{equation}
  P_{ij} = \iint\limits_{\mathcal{D}_i} f_j(t_0,t_1) dt_0 dt_1
  \label{eq:genIntegral}
\end{equation}
where $\mathcal{D}_i$ represents the integration limits associated with $\cH{i}$. A detailed expansion of~\eqref{eq:genIntegral} for all classes is presented in Appendix~\ref{app:PronIntegrals}. 
The integral \eqref{eq:genIntegral} was implemented using MATLAB$^{\copyright}$ function \emph{integral2.m}.

\begin{figure*}[bth]
 \centering
\includegraphics[width=\textwidth]{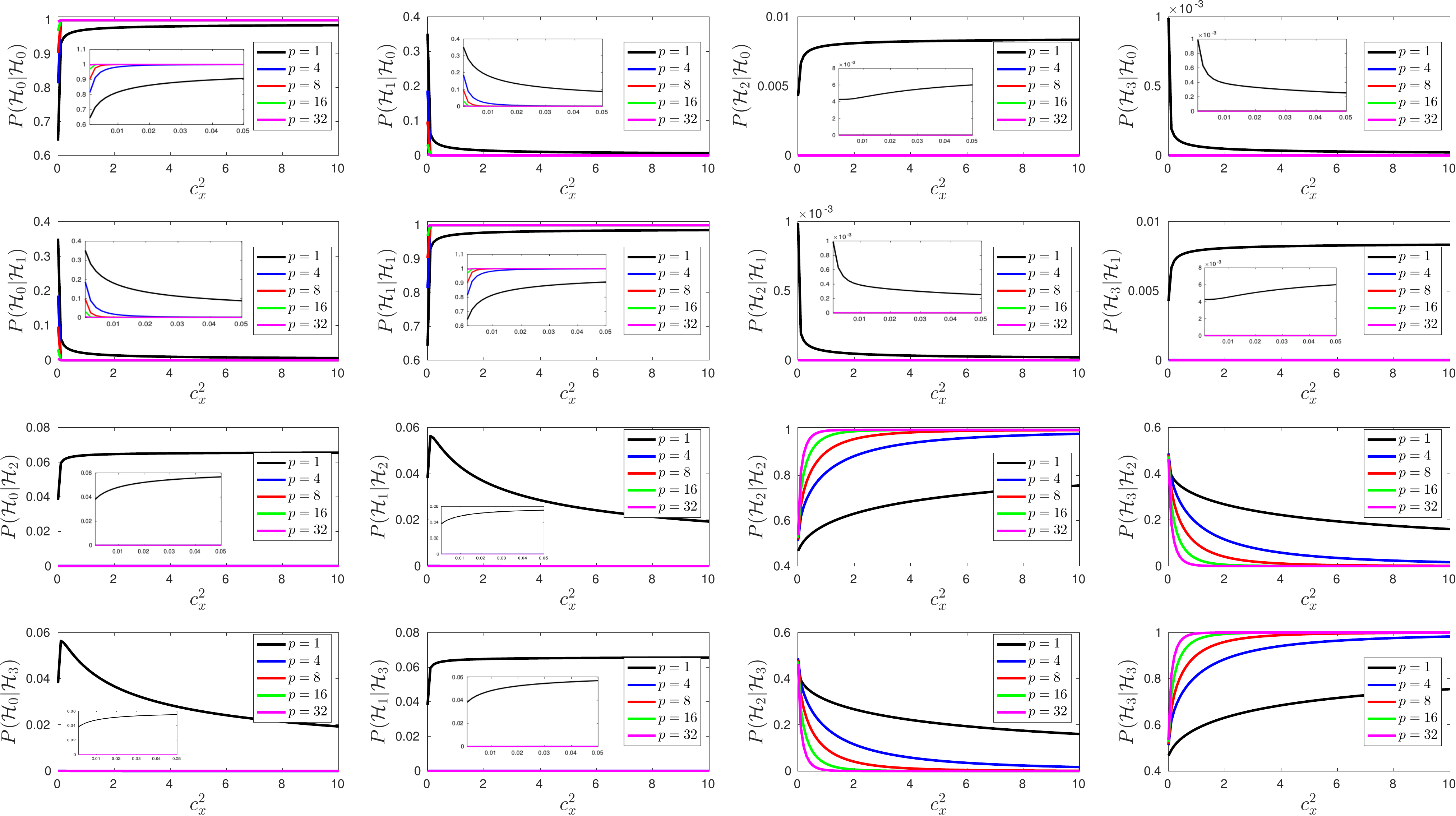}
 \caption{Theoretical performance curves for single- and multi-sample cases ($\sigma_1^2=1$, $\sigma_0^2=0.001$).}
 \label{fig:ms_theo_performance}
\end{figure*}

\begin{figure*}[bth]
 \centering
  \includegraphics[width=\textwidth]{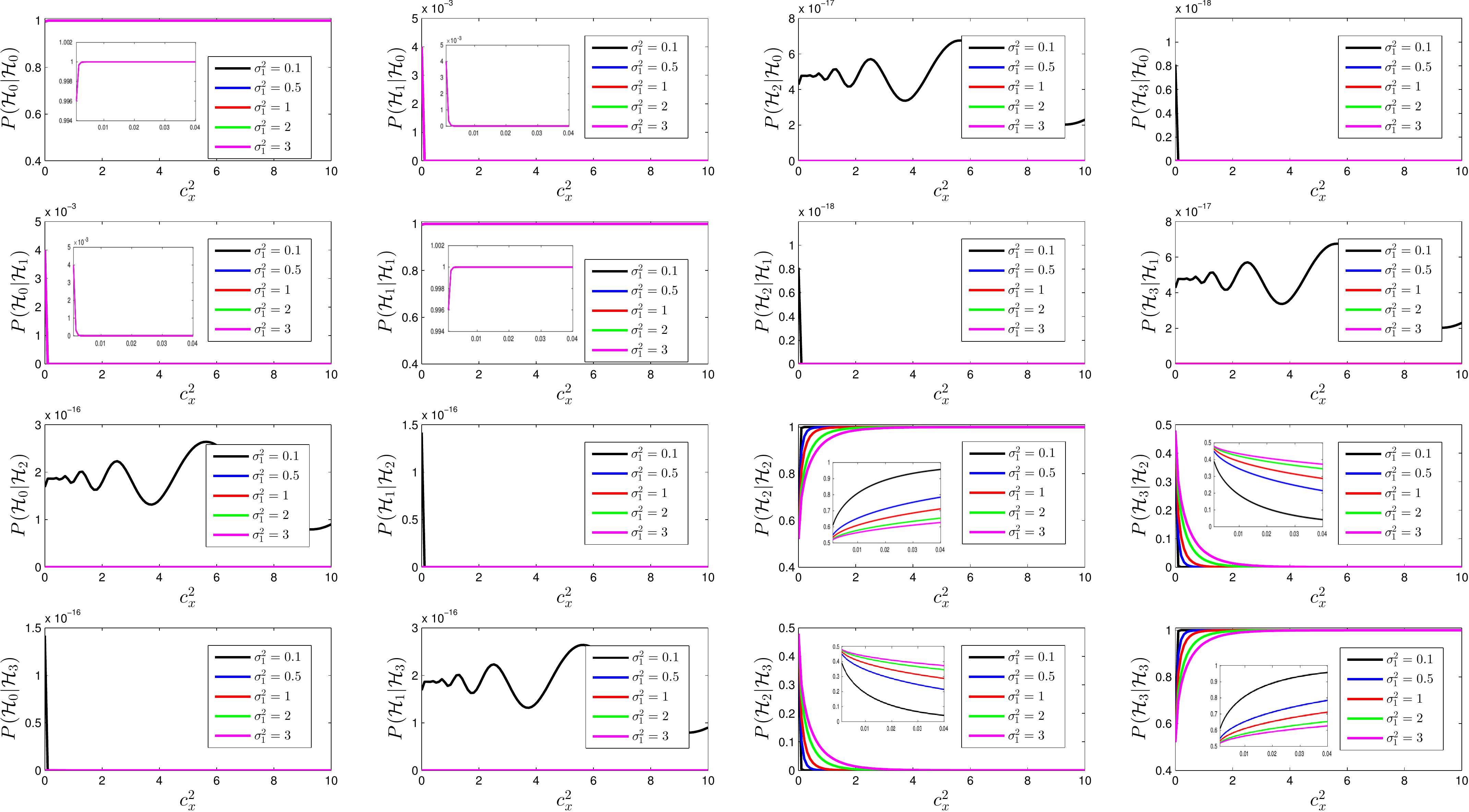}
 \caption{Theoretical performance curves for different values of DT power ($p=32$, $\sigma_0^2=0.001$).}
 \label{fig:ms_theo_performanceDiffDTPw}
\end{figure*}

\begin{figure*}[bth]
 \centering
 \includegraphics[width=\textwidth]{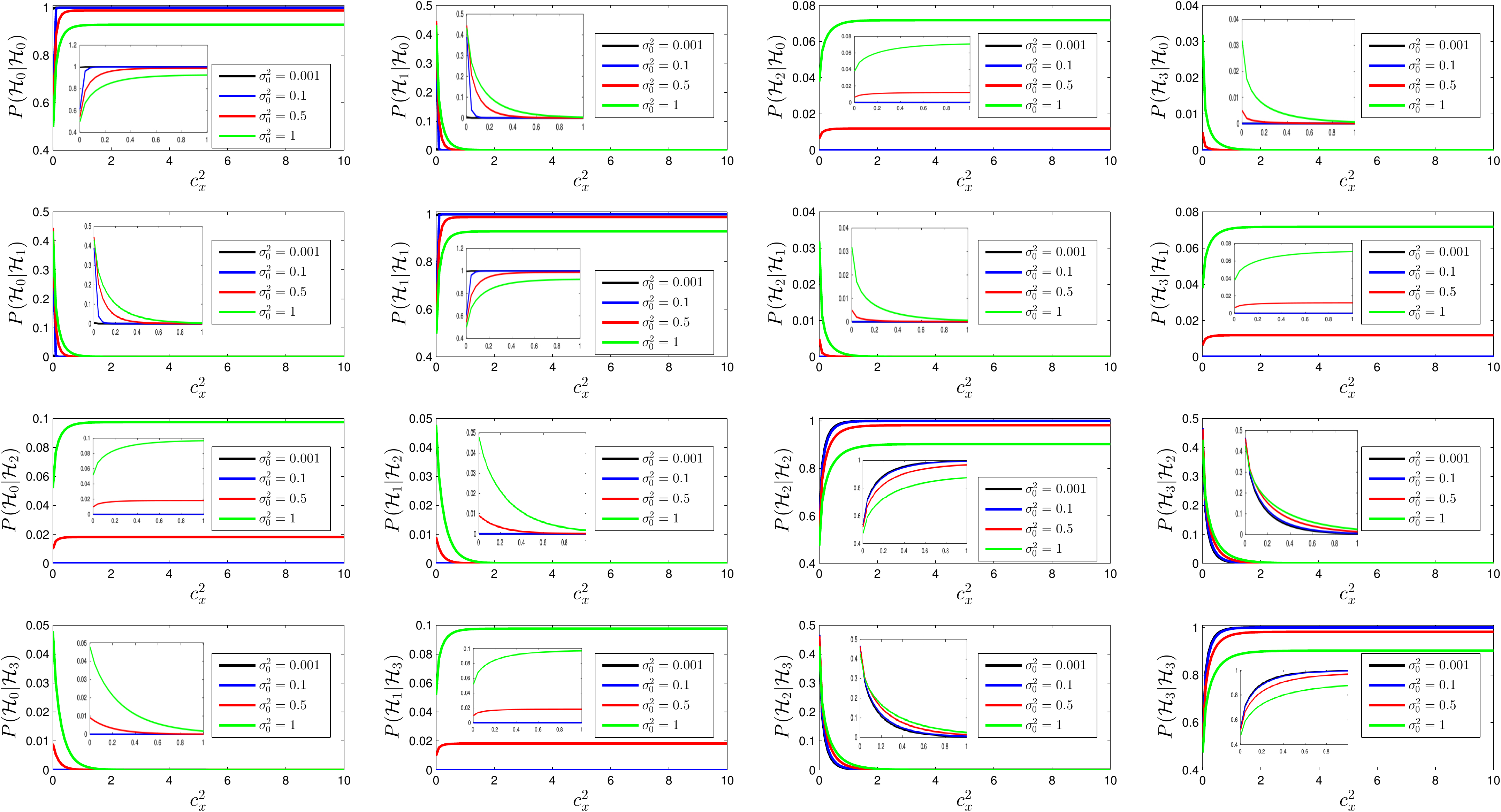}
 \caption{Theoretical performance curves for different values of noise power ($p=32$, $\sigma_1^2=1$).}
 \label{fig:ms_theo_performanceDiffNoisePw}
\end{figure*}

Figures~\ref{fig:ms_theo_performance}--\ref{fig:ms_theo_performanceDiffNoisePw} show the probabilities $P(\mathcal{H}_i|\mathcal{H}_j)$  computed using~\eqref{eq:genIntegral} as functions of $c_x^2 \in [0,10]$ for different sets of parameters. Each row of these figures corresponds to a given true hypothesis $\cH{i}$, $i=0,\ldots,3$. Figure~\ref{fig:ms_theo_performance} shows $P(\mathcal{H}_i|\mathcal{H}_j)$ for $\sigma_1^2=1$, $\sigma_0^2=0.001$, and $p \in\{1, 4, 8, 16, 32\}$. These plots clearly show that the performance of the classifier improves by increasing $c_x^2$ or $p$. A large value of $p$ is especially important in distinguishing between hypotheses $\cH{2}$ and $\cH{3}$.  It is also clear that the classification error increases significantly for low values of $c_x^2$. As a limiting situation, the vector $\vd(n)$ will be placed exactly on the line $\|\vz_0(n)\|^2 = \|\vz_1(n)\|^2$ separating the classes $\cH{0}$ and $\cH{1}$, or $\cH{2}$ and $\cH{3}$ (see Fig.~\ref{fig:Decision}) for $c_x^2=0$.

\begin{figure*}[bth]
\centering
\includegraphics[width=\textwidth]{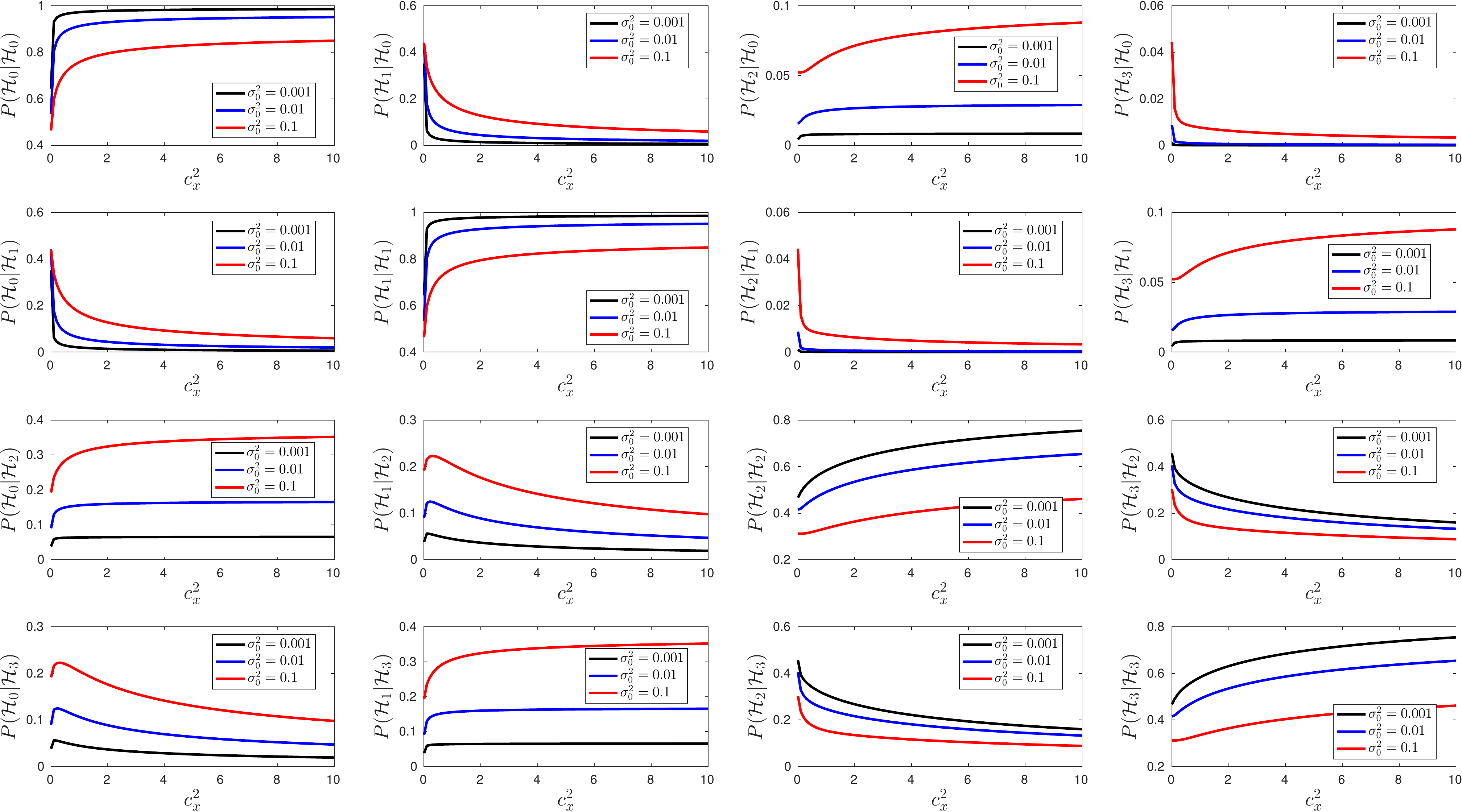}
\caption{One-sample case, different $\sigma_0^2$ values, $P(\mathcal{H}_i|\mathcal{H}_j)$.}
\label{fig:oneSampleTheo}
\end{figure*}

\begin{figure*}[bth]
\centering
\includegraphics[width=\textwidth]{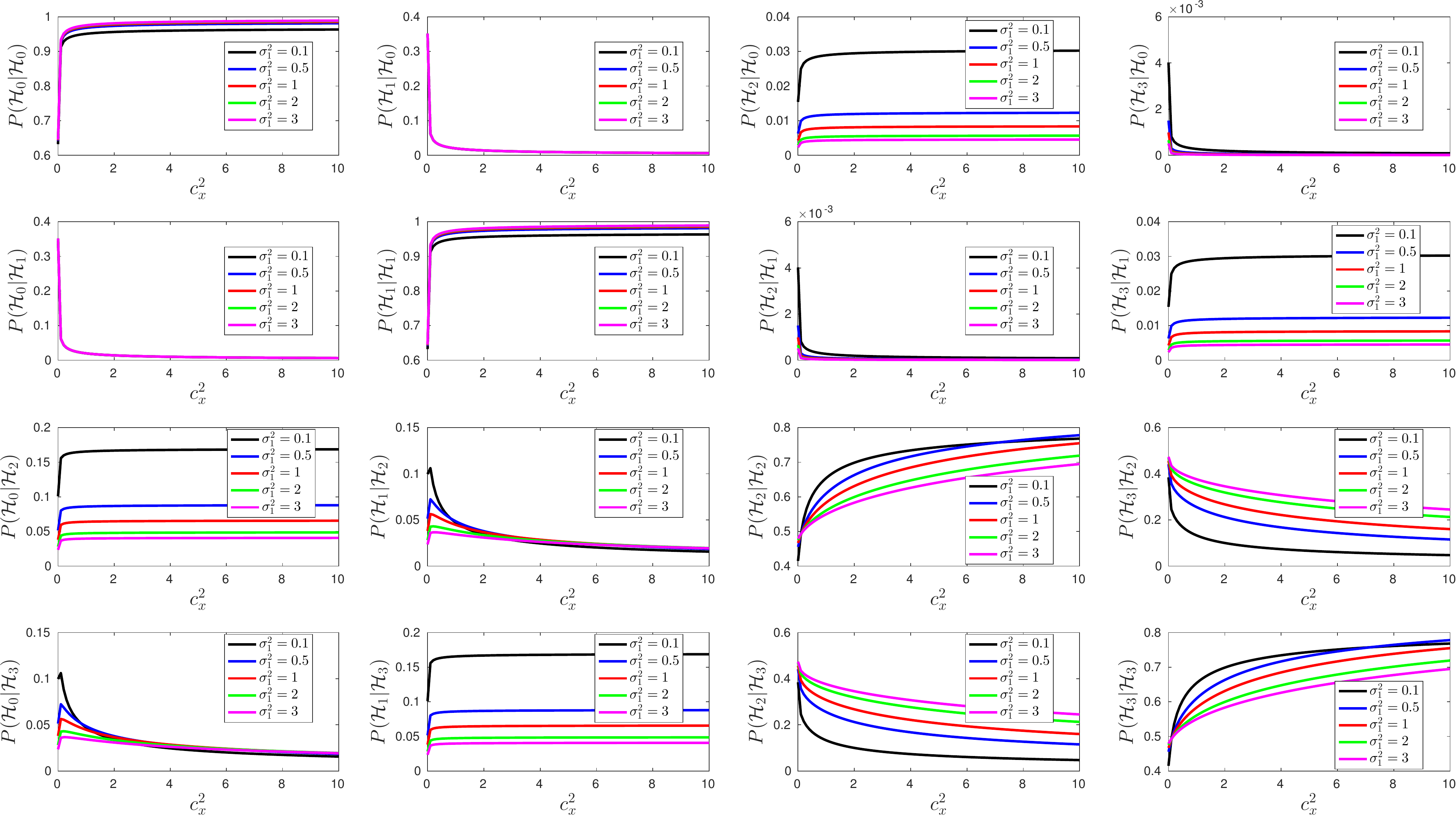}
\caption{One-sample case, different $\sigma_1^2$ values, $P(\mathcal{H}_i|\mathcal{H}_j)$.}
\label{fig:oneSampleTheoDiffDT}
\end{figure*}

Since $p=32$ yielded good classification performance, we opted for fixing $p=32$ in Figs.~\ref{fig:ms_theo_performanceDiffDTPw} and~\ref{fig:ms_theo_performanceDiffNoisePw}, while varying the DT power in Fig.~\ref{fig:ms_theo_performanceDiffDTPw} and the noise power in Fig.~\ref{fig:ms_theo_performanceDiffNoisePw}. Although the DT power has little influence on the classifier performance under $\cH{0}$ and $\cH{1}$ hypotheses (Fig.~\ref{fig:ms_theo_performanceDiffDTPw}), a clearer influence is observable under $\cH{2}$ and $\cH{3}$. In this case, increasing the DT power tends to increase $P(\cH2|\cH3)$ and $P(\cH3|\cH2)$ (bottom two rows of Fig.~\ref{fig:ms_theo_performanceDiffDTPw}). This behavior is expected as the effect of a channel change in distinguishing between hypotheses $\cH2$ and $\cH3$ diminishes with the increase of DT power. Figure~\ref{fig:ms_theo_performanceDiffNoisePw} explores the effect of the noise power on the classifier performance. It can be noted that a large noise power increases the probability of error in detecting the onset of DT ($P(\cH2|\cH0)$, $P(\cH3|\cH1)$, $P(\cH0|\cH2)$, $P(\cH1|\cH3)$), as the performance is a function of the DT to noise ratio $\sigma_1^2/\sigma_0^2$. This effect, however, is very small for ratios larger than 3dB, which is typical in practice. Simulations for the one-sample case with different noise and DT powers are available in Figs.~\ref{fig:oneSampleTheo} and~\ref{fig:oneSampleTheoDiffDT} respectively. Although the results obtained for the one-sample case show (as expected) a stronger influence of DT and noise power in the classification performance when compared to the results for $p=32$,  they corroborate the above conclusions.

\section{Monte Carlo Simulations} \label{sec:MC_Sim}

In this section Monte Carlo (MC) simulations are performed and compared with the theoretical expressions derived in the previous section. These results are also valuable to assess the effect of the independence approximation on the analysis accuracy.

To generate the statistics $\vd(n)$ by sampling the $(2p)$-dimensional vectors $\vz_{2d}(n)$ from $\mathcal{N}(\bb 0, \bb\Sigma_{i2})$, we need to define the covariance ($\bb{\Sigma}_x$) and  correlation ($\bb{R}_{k,x}$) matrices.
Considering the input signal to be auto-regressive of order 1 (AR-1), $\bb{\Sigma}_x$ was chosen as follows~\cite{BershadTourneret2006}:

\begin{equation} \label{eq:sigx}
\boldsymbol{\Sigma}_{x}= \sigma_x^2\left(
\begin{array}{cccc}
1 & \rho & \cdots & \rho^{N-1} \\
\rho & 1 & \cdots & \rho^{N-2} \\
\vdots & \vdots & \ddots & \vdots\\
\rho^{N-1} & \rho^{N-2} & \cdots & 1 \\
\end{array}
\right)
\end{equation}
where $\rho$ controls the input signal correlation.
Thus, the entries of $\bb{R}_{k,x}= E\left[\vx(n)\vx\ttop(n-k)\right]$ can be written as
\begin{equation}
\left[\bb{R}_{k,x}\right]_{ij}=\sigma_x^2\rho^{|i-j-k|}.
\end{equation}
Note that by fixing the vectors $\vh_0$ and $\vh_1$, $\bb{H}_x$ depends only on $\sigma_x^2$, and $\rho$. Thus, for a given $c_x^2$, $\sigma_x^2$ can be easily computed using~\eqref{eq:sigx} and~\eqref{eq:cx}. The vectors $\vh_0$ and $\vh_1$ were assumed to have 1024 samples, and were constructed using the one-sided exponential channels (see~\cite{BershadTourneret2006} and~\cite{TourneretBershadBermudez2009})
\begin{equation}
\vh_i(k) = \left\{
\begin{array}{ll}
c(0.95)^{k-\Delta_i},& k\geq \Delta_i\\
0,& \text{otherwise}
\end{array}
\right.
\end{equation}
where $\Delta_i$ is a relative delay of the channel $\vh_i$ and the parameter $c$ is defined by the filter gain $G = \vh_0^\top \vh_0 = \vh_1^\top \vh_1$. Two different scenarios are studied here corresponding to $G=-10$dB (electrical application) and $G=6$dB (acoustic application).

\begin{figure*}[bth]
\centering
\includegraphics[width=\textwidth]{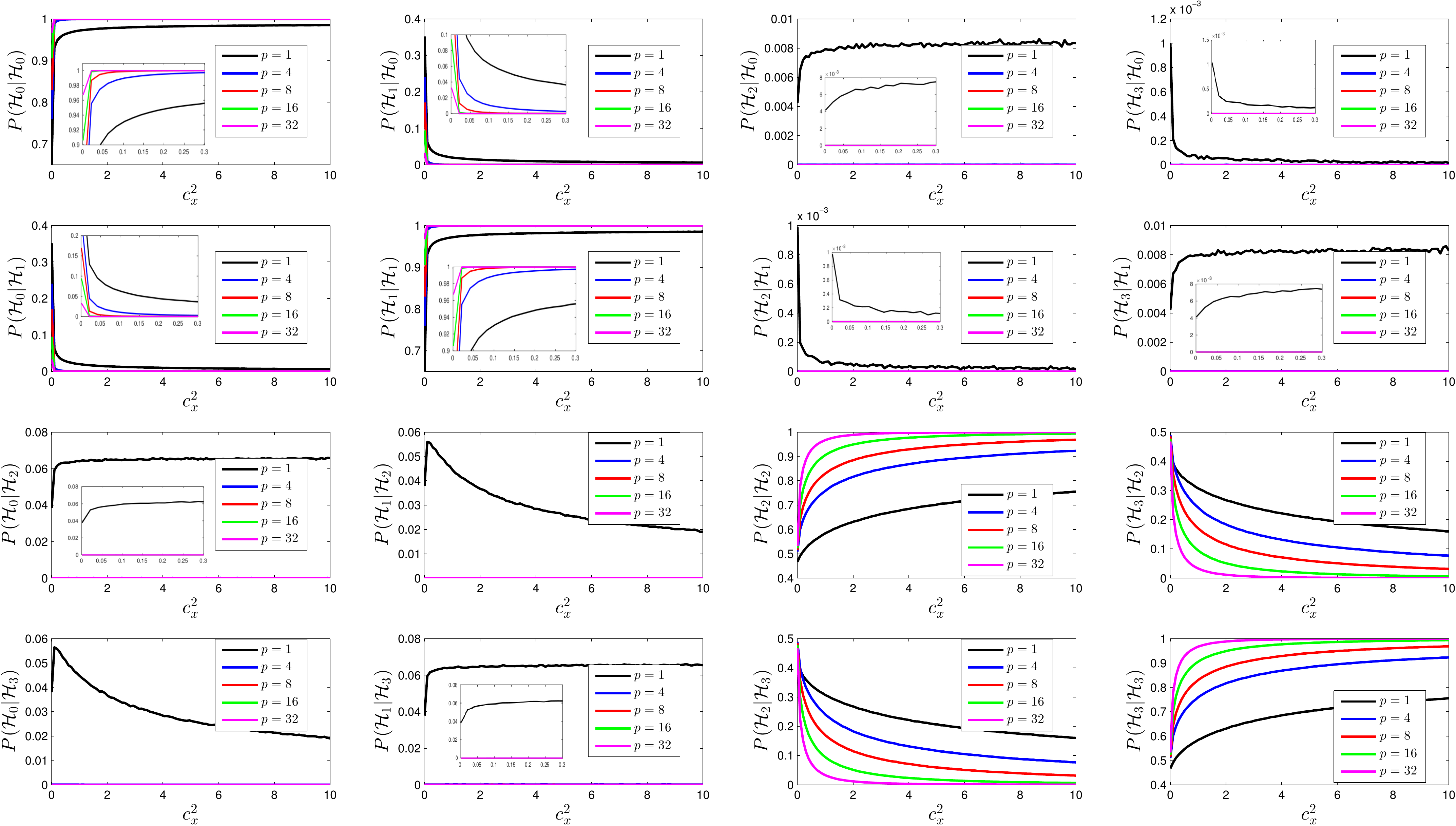}
\caption{MC performance curves assuming AR-1 input signal, $\vz_{pd}(n)$ sampled from $\mathcal{N}(\bb 0,\bb\Sigma_{ip})$, $G=-10$dB (electric application), $\sigma_1^2=1$, $\sigma_0^2=0.001$, $\rho = 0.5$.}
\label{fig:mc_ar1_g1}
\end{figure*}

\begin{figure*}[bth]
\centering
\includegraphics[width=\textwidth]{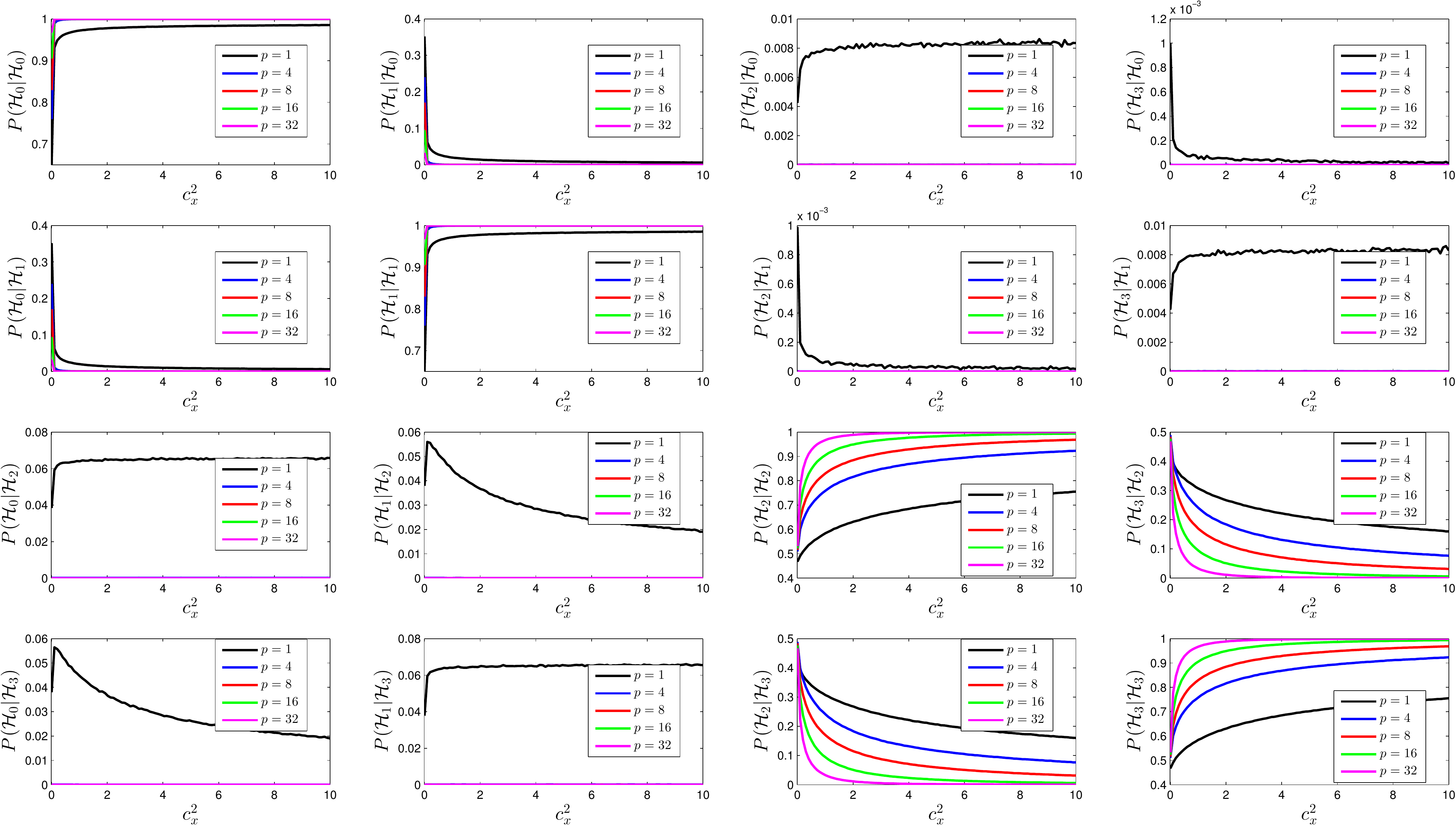}
\caption{MC performance curves assuming AR-1 input signal, $\vz_{pd}(n)$ sampled from $\mathcal{N}(\bb 0,\bb\Sigma_{ip})$, $G=6$dB (acoustic application), $\sigma_1^2=1$, $\sigma_0^2=0.001$, $\rho = 0.5$.}
\label{fig:mc_ar1_g4}
\end{figure*}

Figure~\ref{fig:mc_ar1_g1} presents the MC simulations obtained by averaging $10^6$ runs for $G=-10$dB, with $c_x^2$ varied in the range $[0,10]$, $\rho=0.5$, $\sigma_1^2=1$, and $\sigma_0^2=0.001$, leading to an SNR of 30dB. 
Figure~\ref{fig:mc_ar1_g4} presents the same simulation for $G=6$dB.
When comparing Figs.~\ref{fig:mc_ar1_g1} and~\ref{fig:mc_ar1_g4} with theoretical results (Fig.~\ref{fig:ms_theo_performance}), only a very small degradation in classification accuracy is noted, mainly for $\cH{2}$ and $\cH{3}$, and $p>1$. This small difference is attributed to the use of the independence approximation. 


\begin{figure*}[bth]
\centering
\includegraphics[width=\textwidth]{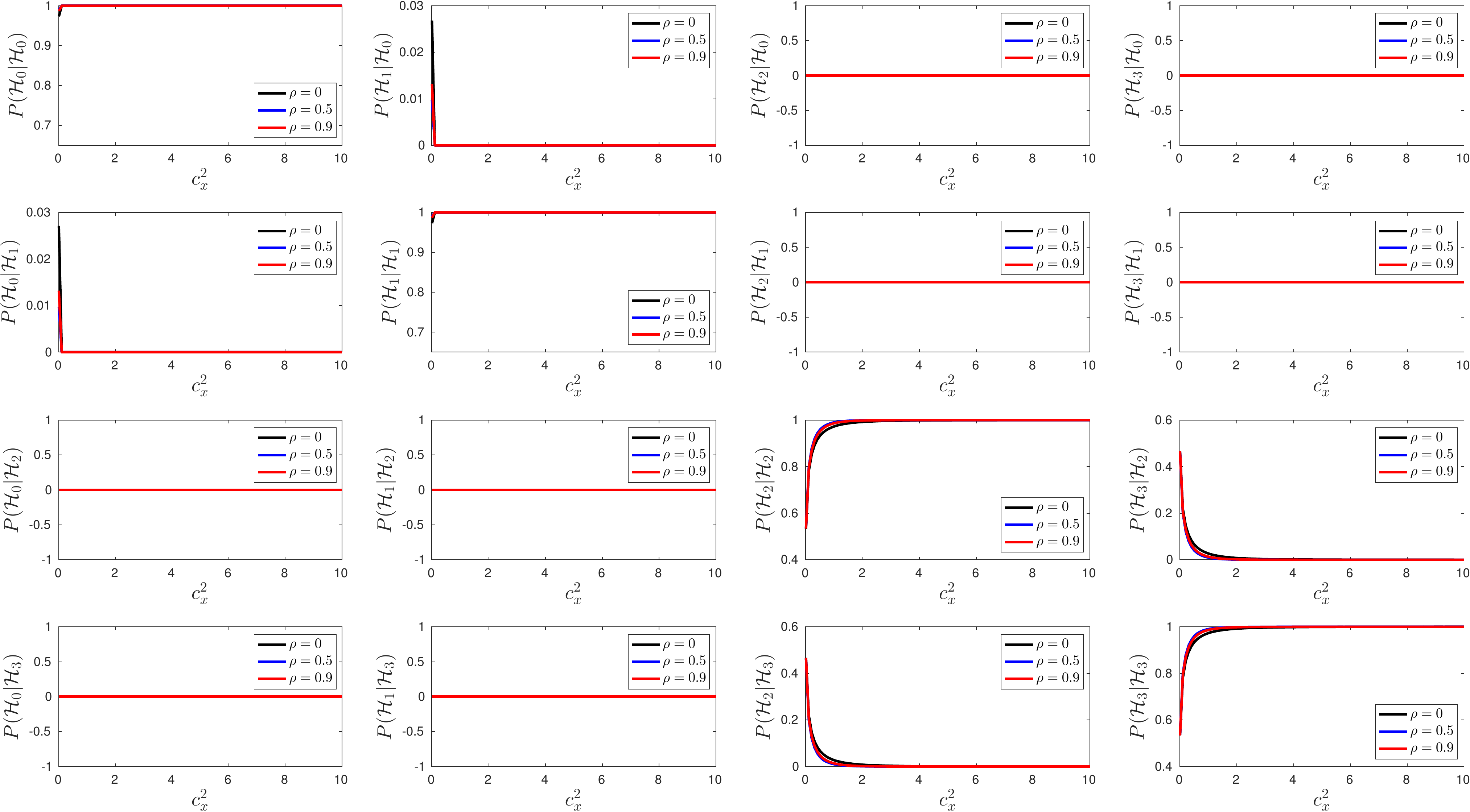}
\caption{MC performance curves assuming AR-1 input signal, $\sigma_x^2\in [0,1]$, $\vz_{pd}(n)$ sampled from $\mathcal{N}(\bb 0,\bb\Sigma_{ip})$, $p=32$, $G=-10$dB (electric application), $\sigma_1^2=1$, $\sigma_0^2=0.001$, $\rho = \{0.0,\,0.5,\,0.9\}$.}
\label{fig:mc_ar1_g1_diffRhos}
\end{figure*}

\begin{figure*}[bth]
\centering
\includegraphics[width=\textwidth]{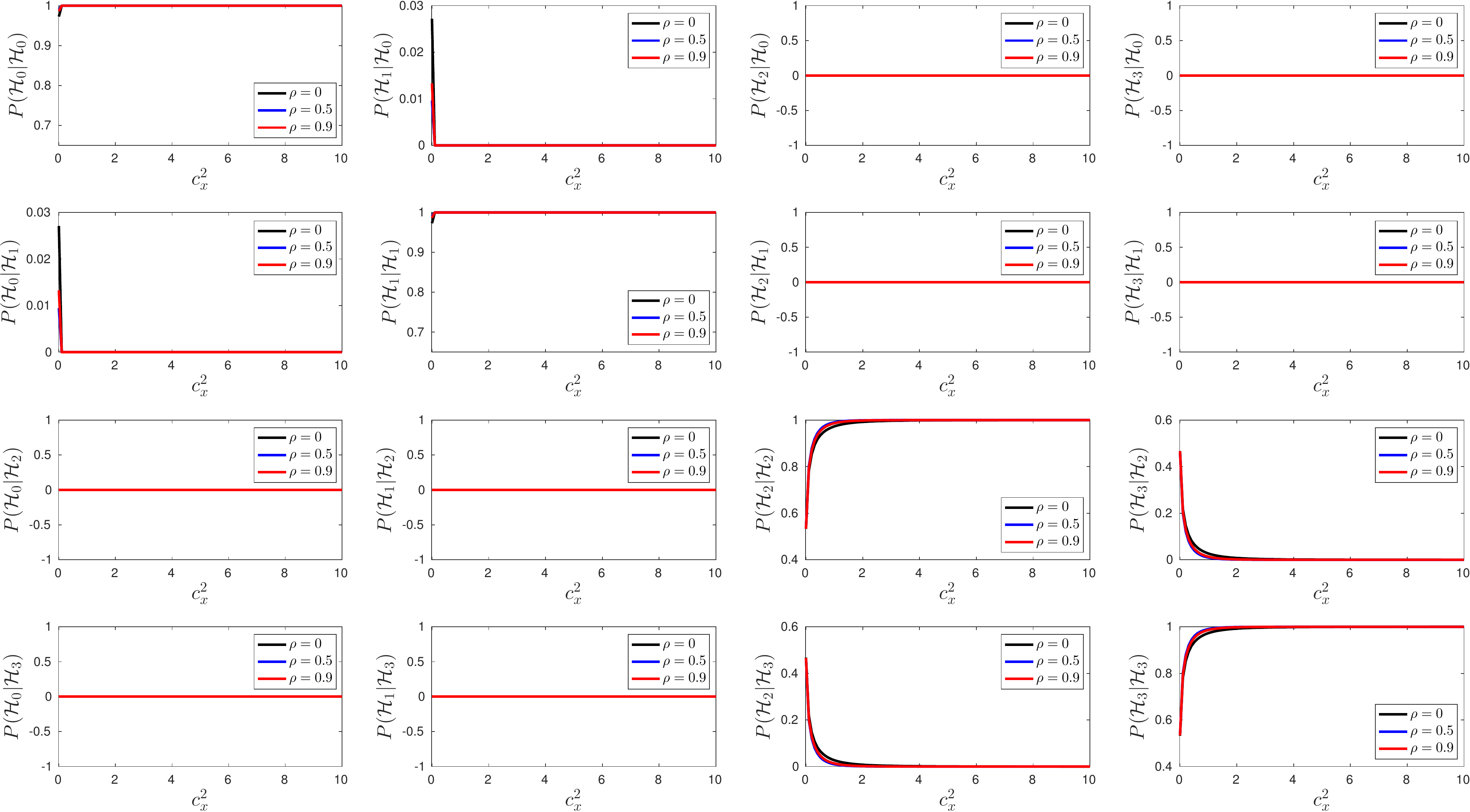}
\caption{MC performance curves assuming AR-1 input signal, $\sigma_x^2\in [0,1]$, $\vz_{pd}(n)$ sampled from $\mathcal{N}(\bb 0,\bb\Sigma_{ip})$, $p=32$, $G=6$dB (electric application), $\sigma_1^2=1$, $\sigma_0^2=0.001$, $\rho = \{0.0,\,0.5,\,0.9\}$.}
\label{fig:mc_ar1_g4_diffRhos}
\end{figure*}

MC simulations for different values of the correlation coefficient $\rho$ are shown in Figs.~\ref{fig:mc_ar1_g1_diffRhos} ($G=-10$dB) and~\ref{fig:mc_ar1_g4_diffRhos} ($G=6$dB). Although varying $\rho$ has little impact on the classification performance, it is interesting to notice that increasing $\rho$ slightly improves the classification performance in all classes, but especially for $\cH{2}$ and $\cH{3}$. 

\section{Application to echo cancellers} \label{sec:section5}

\subsection{Control strategy}\label{sec:controlStrat}
The classification hypotheses presented in~\eqref{Neyman2} considered that in each case the adaptive filter had time to converge or diverge. This becomes a critical point for designing the control block (see, Fig.~\ref{fig:Echo}) since the probabilities of error are high for low values of $c_x^2$. Two direct consequences related to this characteristic are the following:
\begin{enumerate}
\item ($\cH{0}/\cH{1}$): Whenever $\vh_0$ is copied to $\vh_1$ $c_x^2$ becomes zero and the probability of error becomes large between classes $\cH{0}$ and $\cH{1}$. In fact, if $\vh_0=\vh_1$ the vector $\vd(n)$ will be exactly in the frontier between the two classes (see, Fig.~\ref{fig:Decision}).

\item ($\cH{1}/\cH{2},\cH{3}$): When CC happens, $\vh_0$ and $\vh_1$ may assume values very far from the new true filter response $\vh_\text{new}$. If this is the case, classification errors ($\cH{2}|\cH{1}$ or $\cH{3}|\cH{1}$) are expected since both norms $\|\vz_i(n)\|^2$, $i = 1,2$, may become larger than $T$. 
\end{enumerate}
To address these problems, we propose a control strategy that combines tuning of the adaptive step size $\mu$, defining an appropriate frequency for the realization of the tests, and introducing a delay before actually changing the system state after each decision.\\

\newpage
\noindent\textbf{Adaptation step}

  The shadow filter $\vh_0$ is always adapting, even during DT, since the difference between $\vh_0$ and $\vh_1$ is crucial for improving classification rates. However, different adaptation step sizes can adopted for each class:
  \begin{enumerate}
  \item During $\cH{0}$, $\mu=\mu_0$ should be low since the aim is to make a fine tuning of the filter coefficients.
  \item During $\cH{1}$, $\mu=\mu_1$ should be set as high as possible to speed-up convergence of the adaptive algorithm.
  \item During $\cH{2}$, $\mu=\mu_2$ should be set to a small value so that $\vh_0$ can diverge slowly under DT, start to converge once DT is over or in the accurence of CC.
  \item Class $\cH{3}$ is critical since it corresponds to the occurence of CC with or without DT signal. Our practical experience indicates that setting $\mu=\mu_3$ to a value between $\mu_0$ and $\mu_1$ leads to good classification results.
\end{enumerate}

\noindent\textbf{Frequency of tests}

The difference filter $\vh_0-\vh_1$ plays a central role in classification accuracy. Hence, it is advisable to allow a minimum number $N_t$ of samples between two tests to allow a clear differenciation of the two responses.\\

\noindent\textbf{Filter copy}

Whenever classes $\cH{0}$ or $\cH{1}$ are detected, the shadow filter $\vh_0$ should be copied to $\vh_1$ if $\|\vz_0(n)\|^2 < \|\vz_1(n)\|^2$. To account for transients occurring after the exit of a given state (especially when DT stops), it is advisable to consider a delay of $N_c<N_t$ samples between the decision moment and the actual filter copy.\\

\noindent\textbf{Decisions in the neighborhood of} $\|\vz_0(n)\|^2 = \|\vz_1(n)\|^2$

Decision between $\cH{0}$ and $\cH{1}$, and between $\cH{2}$ and $\cH{3}$ are rather arbitrary in practical situations when $\|\vz_0(n)\|^2 \approx \|\vz_1(n)\|^2$. To address this issue, we propose to allow changes between classes $\cH{0}$ and $\cH{1}$, or between $\cH{2}$ and $\cH{3}$ only if  
$$1-\varepsilon\leq \frac{\|\vz_0(n)\|^2}{\|\vz_1(n)\|^2} \leq 1+\varepsilon$$
where $\varepsilon\in[0,1)$.

\subsection{Synthetic Data}\label{sec:synthData}
\begin{figure}[bth]
\centering
\includegraphics[width=0.48\textwidth]{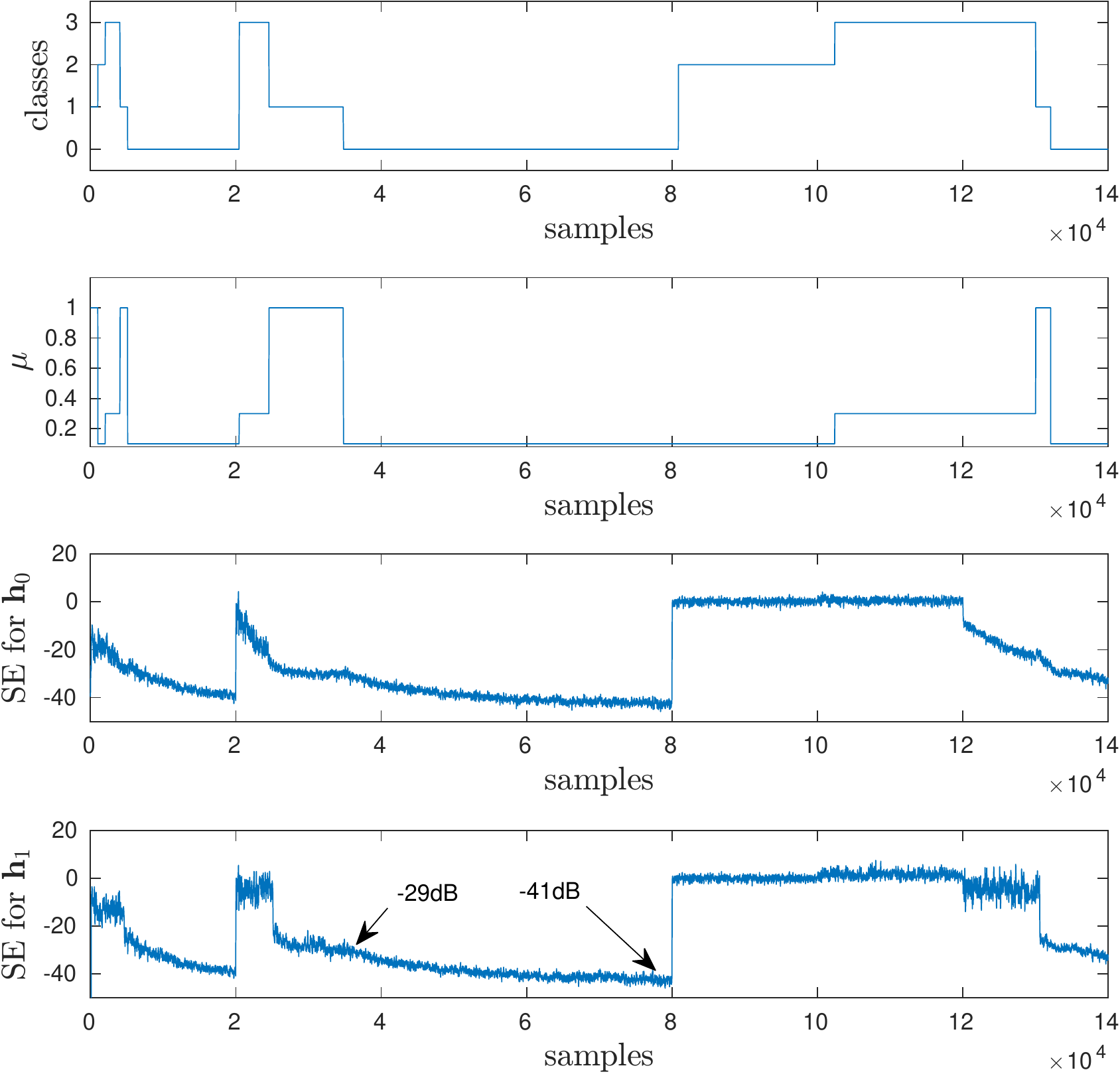}
\caption{Performance of the echo canceller system ($G=-10$dB). From top down, the panels present the evolution of the classification result (top), adaptation step size $\mu$, SE in dB for $\vh_0$ and $\vh_1$ (bottom). $\sigma_1^2=1$, $\sigma_0^2=0.001$, $\varepsilon=0.25$, $N_t=1024$, $N_c=512$. }
\label{fig:synthData_G1}
\end{figure}

This section considers the AR-1 ($\rho=0.5$) data discussed in Section~\ref{sec:MC_Sim}, and also used in~\cite{BershadTourneret2006,TourneretBershadBermudez2009}. We considered filter responses $\vh_0$ and $\vh_1$ with $N=1024$ samples, and fixed the parameters $p=32$, $\sigma_0^2=0.001$, and $\sigma_1^2=1$. The signal $y(n)$ consisting of $140K$ samples ($K=1000$) was formulated as
\begin{equation}
y(n) = \left\{
\begin{array}{ll}
\vh_0^\top\vx(n) + n_0(n),& n\in\mathcal{I}_1\\
\vh_1^\top\vx(n) + n_0(n),& n\in\mathcal{I}_2\\
\vh_1^\top\vx(n) + n_1(n) + n_0(n),& n\in\mathcal{I}_3\\
\vh_2^\top\vx(n) + n_1(n) + n_0(n),& n\in\mathcal{I}_4\\
\vh_2^\top\vx(n) + n_0(n),& n\in\mathcal{I}_5
\end{array}
\right.
\end{equation}
with intervals $\mathcal{I}_1 = [0, 20K]$, $\mathcal{I}_2 = (20K, 80K]$, $\mathcal{I}_3 = (80K, 100K]$, $\mathcal{I}_4= (100K, 120K]$, and $\mathcal{I}_5=(120K, 140K]$. Hence, CC occurs at sample 20,001, DT occurs between samples 80,001 and 120,000, and a second CC occurs during the DT period at sample 100,001. The adaptive algorithm employed was the Normalized Least Mean Square (NLMS) algorithm, whose maximum convergence speed is known to be attained for $\mu=1$~\cite{Haykin:1996:AFT:230061}. The control parameters were set to $N_t=1024$, $N_c=512$, $\mu_0=\mu_2=0.1$, $\mu_1=1$, $\mu_3=0.3$, and $\varepsilon$ was set to 0.25 for $G=-10dB$. The adaptive filter coefficients were initialized equal to zero and the adaptation step was initialized as $\mu = \mu_1$ (CC). The simulation results for one realization of the synthetic signal are shown in Fig.~\ref{fig:synthData_G1}. The top panel presents the classes attributed by the classifier to each sample in time. The second panel presents the step-size corresponding to each class. The squared excess errors (SE) $e_i^2(n)=(y(n)-\vh_i^\top\vx(n) -n_0(n))^2$, $i=\{0,1\}$, for $\vh_0$ and $\vh_1$ follow in the bottom two panels. Although the good classification performance is evident in this example, the $\cH{1}/\cH{2},\cH{3}$ issue discussed in section~\ref{sec:controlStrat} can be noticed after the CC at sample 20K. 
The samples are classified as $\cH{3}$ before $\|\vz_0(n)\|^2$ becomes smaller then $T_p$. Then the correct class $\cH{1}$ is selected before sample 30K. However, since the adaptive filter never stops adapting, this problem is satisfactorily mitigated without severe deterioration of the echo canceler performance, as can be verified by the SE results in the two bottom panels.
%
These results clearly show the performance improvement resulting from the generalization of the approach proposed in~\cite{BershadTourneret2006,TourneretBershadBermudez2009}. The improvement shows especially during the single-talk periods. As DT or CC do not occur during these periods, the proposed solution leads to a reduction of the step size $\mu$, clearly improving the quality of channel estimation. Note, for instance, that the step size reduction that happens at iteration 35K due to the acceptance of hypothesis $\cH{0}$ leads to a drop in SE that reaches 12~dB at iteration 80K.

\begin{figure}[bth]
\centering
\includegraphics[width=0.48\textwidth]{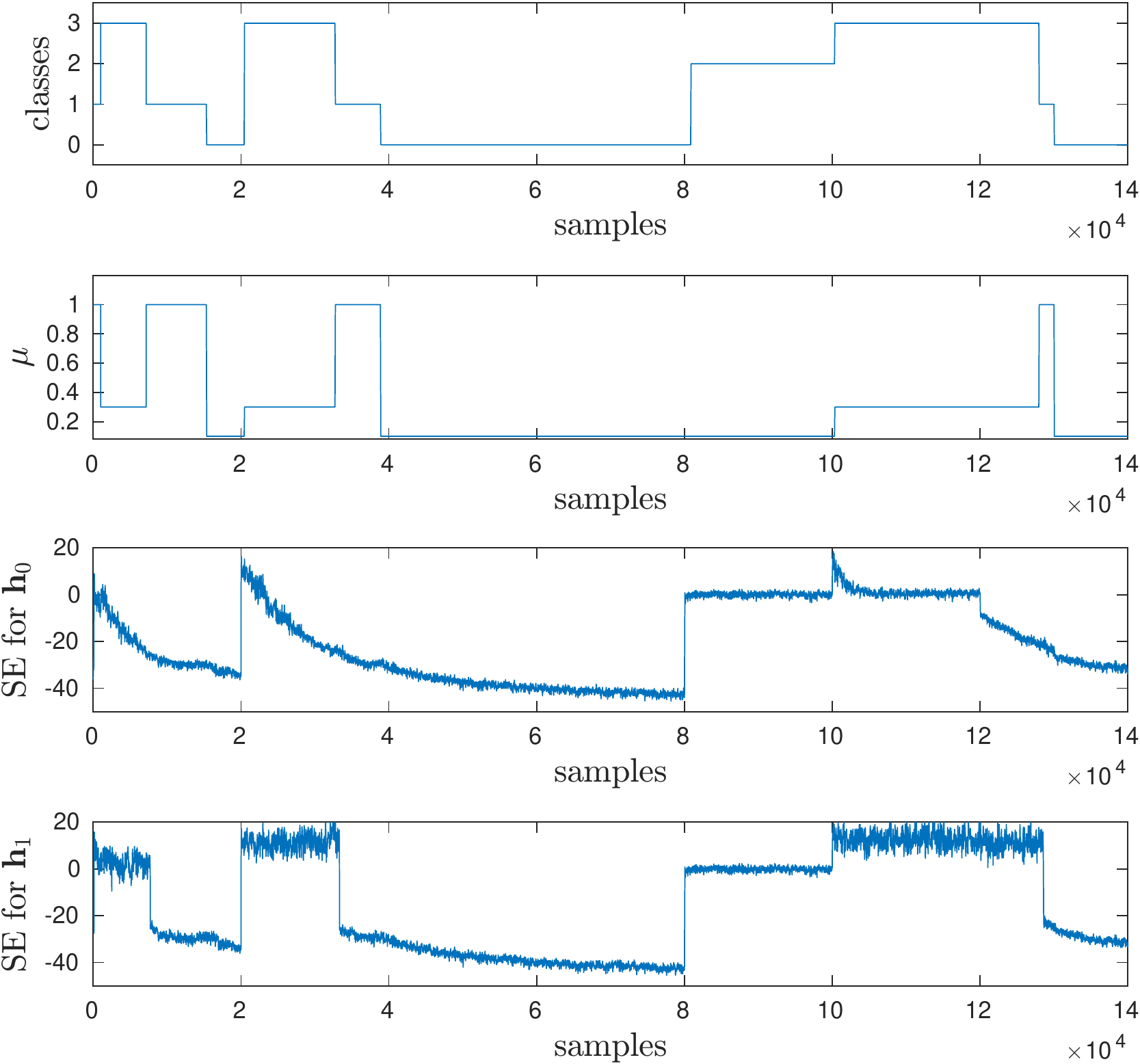}
\caption{Performance of the echo canceller system ($G=6$dB). From top down, the panels present the evolution of the classification result (top), adaptation step size $\mu$, MSE in dB for $\vh_0$ and $\vh_1$ (bottom). $\sigma_1^2=1$, $\sigma_0^2=0.001$, $\varepsilon=0.3$, $N_t=1024$. }
\label{fig:synthData_G4}
\end{figure}

Simulations with $G=6dB$ yielded similar results, and are displayed in Fig.~\ref{fig:synthData_G4}.

\subsection{Voice Data Over a Real Channel}
\begin{figure}[bth]
\centering
\includegraphics[width=0.48\textwidth]{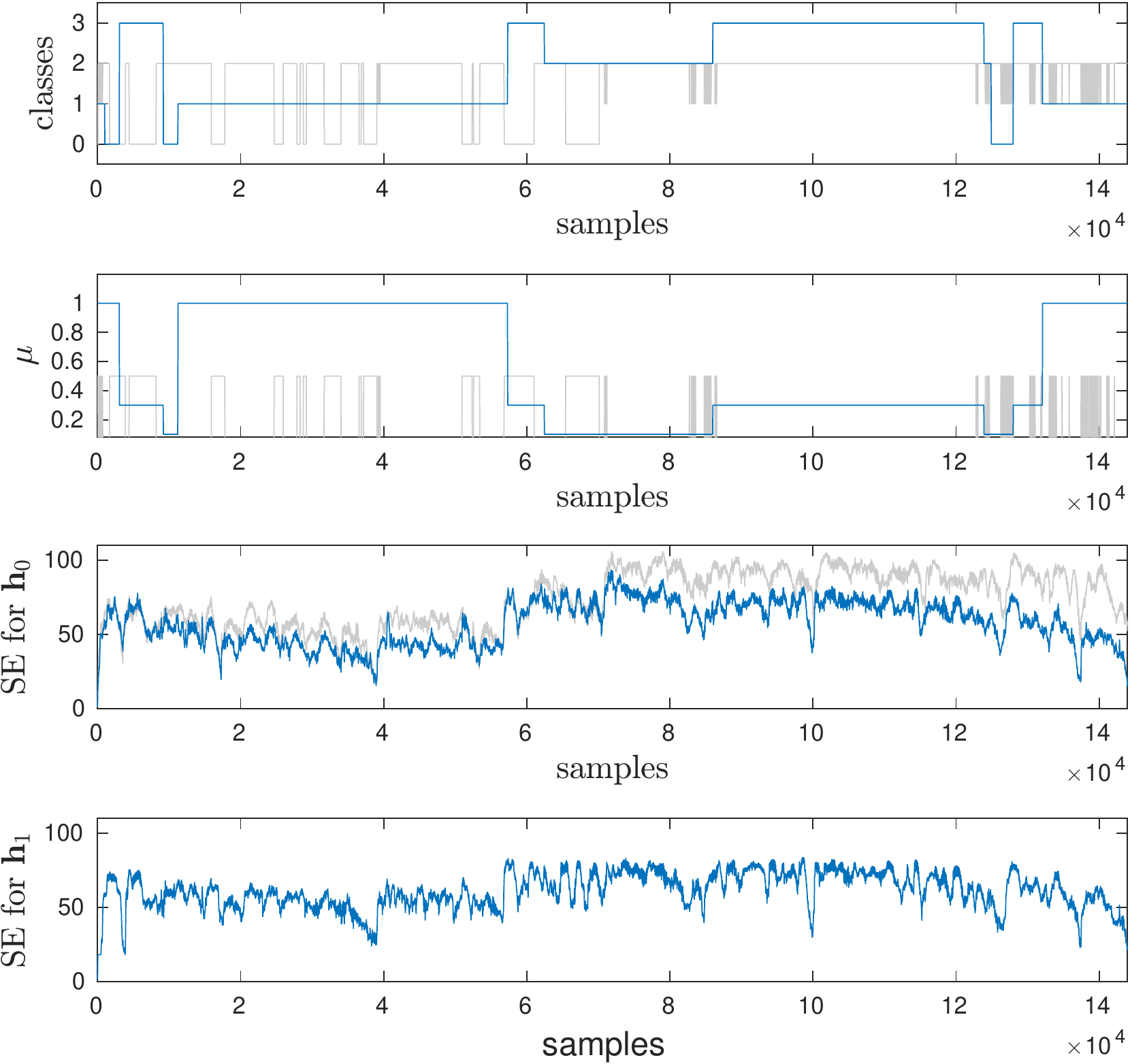}
\caption{Performance of the echo canceller system for voice over real channels. From top down, the panels present the evolution of the classification result (top), adaptation step size $\mu$, SE in dB for $\vh_0$ and $\vh_1$ (bottom). $\varepsilon=0.25$, $N_t=1024$, $N_c=512$. Results for the proposed method (blue) and using the method in~\cite{jung2005} (gray).}
\label{fig:realData}
\end{figure}

For the simulation presented in this section we used the same voice data and channels considered in~\cite{BershadTourneret2006,TourneretBershadBermudez2009}. The data is approximately 144K samples long, with two CC's occurring at sample 50K and 123K, and an intense DT occurring between 57-123K. The simulation results presented in Fig.~\ref{fig:realData} compare the proposed decision framework (blue) with the sequential classification strategy presented in~\cite{jung2005} (gray). 
To deal with the power fluctuation inherent in speech signals, we used $p=3000$ and set the detection threshold $T_p=1.2\times 10^6$ chosen empirically to avoid $\cH{0}$ and $\cH{1}$ errors during DT. The remaining control strategy parameters were kept the same used in the synthetic simulation presented in Fig.~\ref{fig:synthData_G1}. The parameters used for the method in~\cite{jung2005} were set to the same values used by the authors. 
Although the detector presented in~\cite{jung2005} also considers different classes, the authors did not consider the influence of multiple samples nor used a shadow filter configuration, which clearly impacts the results. 
The results displayed in Fig.~\ref{fig:realData} can be also compared with the result obtained in~\cite[Fig. 9]{TourneretBershadBermudez2009}, which indicates that the proposed classification and control strategies perform at least as well as previous echo cancellation systems.


\section{Results and Conclusions} \label{section6}
In this manuscript we presented a low computational cost multi-class classifier with a coupling control strategy for the echo cancellation problem. 
The proposed classification rule initially proposed for one-sample was easily extended to the multi-sample scenario. Error probabilities were also analytically computed under the assumption of independence among vectors $\vz(n-k)$. This assumption led to bivariate gamma distributions for the sufficient statistics $\vd(n)$ and performance curves that proved accurate when confronted with Monte Carlo Simulations. The results showed that the greater flexibility provided by the multi-class approach could be well explored by the control strategy which considered different step-sizes under each hypothesis. The simulations with synthetic data showed that the multi-class strategy is viable if accurate double-talk and noise power can be estimated, improving the filter convergence during long periods of single-talk. Simulations in a more realistic scenario (voice over real channels) showed that the proposed strategy works as well as other methods in the literature even ignoring the power fluctuation of speech signals and using a fixed threshold $T_p$.

\appendices
\section{Classification rule} \label{ClassificationRule}
This appendix derives the classification rule \eqref{simple} for the
one sample case. This rule corresponds to accepting hypothesis
$\mathcal{H}_i$ if
\begin{equation} \label{logclassif2}
\boldsymbol{z}\ttop(n) \left( \boldsymbol{\Sigma}_{j1}^{-1} -
\boldsymbol{\Sigma}_{i1}^{-1} \right) \boldsymbol{z}(n)  \ge \ln
\left( \frac{|\mS_{i1}|}{|\mS_{j1}|}\right)
\end{equation}
for all $j \neq i$. As a consequence
hypothesis $\mathcal{H}_0$ is accepted if the three following
conditions are satisfied
\begin{equation*}
\begin{split}
&\vz\ttop(n) \left( \boldsymbol{\Sigma}_{11}^{-1} -
\boldsymbol{\Sigma}_{01}^{-1} \right) \vz(n)  \ge \ln
\left( \frac{|\mS_{01}|}{|\mS_{11}|}\right) \\
&\vz\ttop(n) \left( \boldsymbol{\Sigma}_{21}^{-1} -
\boldsymbol{\Sigma}_{01}^{-1} \right) \vz(n)  \ge \ln
\left( \frac{|\mS_{01}|}{|\mS_{21}|}\right) \\
&\vz\ttop(n) \left( \boldsymbol{\Sigma}_{31}^{-1} -
\boldsymbol{\Sigma}_{01}^{-1} \right) \vz(n)  \ge \ln \left(
\frac{|\mS_{01}|}{|\mS_{31}|}\right).
\end{split}
\label{A1}
\end{equation*}
By replacing the matrix inverses and determinants in these
expressions, the following results are obtained
\begin{equation*}
\begin{split}
&z_0^2(n) - z_1^2(n)>0 \\
&z_1^2(n) \left( \frac{1}{\sigma_0^2+\sigma_1^2} -\frac{1}{\sigma_0^2} \right)> \ln \left( 1+ \frac{\sigma_1^2}{\sigma_0^2}\right) \\
&\frac{z_0^2(n)}{\sigma_0^2+\sigma_1^2}-\frac{z_1^2(n)}{\sigma_0^2}
> - \ln \left( 1+ \frac{\sigma_1^2}{\sigma_0^2}\right).
\end{split}
\label{A2}
\end{equation*}
These three conditions are equivalent to
\begin{equation*}
z_1^2(n) < z_0^2(n) \; \textrm{and} \; z_1^2(n) < \frac{\sigma_0^2
(\sigma_0^2+\sigma_1^2)}{\sigma_1^2} \ln \left( 1 +
\frac{\sigma_1^2}{\sigma_0^2} \right).
\end{equation*}

Hypothesis $\mathcal{H}_1$ is accepted if the three following
conditions are satisfied
\begin{equation*}
\begin{split}
&\vz\ttop(n) \left( \boldsymbol{\Sigma}_{01}^{-1} -
\boldsymbol{\Sigma}_{11}^{-1} \right) \vz(n)  \ge \ln
\left( \frac{|\mS_{11}|}{|\mS_{01}|}\right) \\
&\vz\ttop(n) \left( \boldsymbol{\Sigma}_{21}^{-1} -
\boldsymbol{\Sigma}_{11}^{-1} \right) \vz(n)  \ge \ln
\left( \frac{|\mS_{11}|}{|\mS_{21}|}\right) \\
&\vz\ttop(n) \left( \boldsymbol{\Sigma}_{31}^{-1} -
\boldsymbol{\Sigma}_{11}^{-1} \right) \vz(n)  \ge \ln \left(
\frac{|\mS_{11}|}{|\mS_{31}|}\right).
\end{split}
\label{A3}
\end{equation*}
Equivalently
\begin{equation*}
\begin{split}
&z_1^2(n) - z_0^2(n)>0 \\
&\frac{z_1^2(n)}{\sigma_0^2+\sigma_1^2}-\frac{z_0^2(n)}{\sigma_0^2}
> - \ln \left( 1+ \frac{\sigma_1^2}{\sigma_0^2}\right) \\
&z_0^2(n) \left( \frac{1}{\sigma_0^2+\sigma_1^2}
-\frac{1}{\sigma_0^2} \right)> -\ln \left( 1+
\frac{\sigma_1^2}{\sigma_0^2}\right).
\end{split}
\label{A4}
\end{equation*}
These three conditions are equivalent to
\begin{equation*}
z_1^2(n) > z_0^2(n) \;\text{and}\; z_0^2(n) < \frac{\sigma_0^2
(\sigma_0^2+\sigma_1^2)}{\sigma_1^2} \ln \left( 1 +
\frac{\sigma_1^2}{\sigma_0^2} \right).
\end{equation*}

Hypothesis $\mathcal{H}_2$ is accepted if the three following
conditions are satisfied
\begin{equation*}
\begin{split}
&\vz\ttop(n) \left( \boldsymbol{\Sigma}_{01}^{-1} -
\boldsymbol{\Sigma}_{21}^{-1} \right) \vz(n)  \ge \ln
\left( \frac{|\mS_{21}|}{|\mS_{01}|}\right) \\
&\vz\ttop(n) \left( \boldsymbol{\Sigma}_{11}^{-1} -
\boldsymbol{\Sigma}_{21}^{-1} \right) \vz(n)  \ge \ln
\left( \frac{|\mS_{21}|}{|\mS_{11}|}\right) \\
&\vz\ttop(n) \left( \boldsymbol{\Sigma}_{31}^{-1} -
\boldsymbol{\Sigma}_{21}^{-1} \right) \vz(n)  \ge \ln \left(
\frac{|\mS_{21}|}{|\mS_{31}|}\right).
\end{split}
\label{A5}
\end{equation*}
Equivalently
\begin{equation*}
\begin{split}
&z_1^2(n) >\frac{\sigma_0^2 (\sigma_0^2+\sigma_1^2)}{\sigma_1^2} \ln
\left( 1 +\frac{\sigma_1^2}{\sigma_0^2} \right) \\
&\frac{z_1^2(n)}{\sigma_0^2+\sigma_1^2}-\frac{z_0^2(n)}{\sigma_0^2}
<- \ln \left( 1+ \frac{\sigma_1^2}{\sigma_0^2}\right) \\
&z_0^2(n)>z_1^2(n).
\end{split}
\label{A6}
\end{equation*}
These three conditions are equivalent to
\begin{equation*}
z_1^2(n) < z_0^2(n) \;\text{and}\;  z_1^2(n) > \frac{\sigma_0^2
(\sigma_0^2+\sigma_1^2)}{\sigma_1^2} \ln \left( 1 +
\frac{\sigma_1^2}{\sigma_0^2} \right).
\end{equation*}

Hypothesis $\mathcal{H}_3$ is accepted if the three following
conditions are satisfied
\begin{equation*}
\begin{split}
&\vz\ttop(n) \left( \boldsymbol{\Sigma}_{01}^{-1} -
\boldsymbol{\Sigma}_{31}^{-1} \right) \vz(n)  \ge \ln
\left( \frac{|\mS_{31}|}{|\mS_{01}|}\right) \\
&\vz\ttop(n) \left( \boldsymbol{\Sigma}_{11}^{-1} -
\boldsymbol{\Sigma}_{31}^{-1} \right) \vz(n)  \ge \ln
\left( \frac{|\mS_{31}|}{|\mS_{11}|}\right) \\
&\vz\ttop(n) \left( \boldsymbol{\Sigma}_{21}^{-1} -
\boldsymbol{\Sigma}_{31}^{-1} \right) \vz(n)  \ge \ln \left(
\frac{|\mS_{31}|}{|\mS_{21}|}\right).
\end{split}
\label{A7}
\end{equation*}
Equivalently
\begin{equation*}
\begin{split}
&\frac{z_1^2(n)}{\sigma_0^2}-\frac{z_0^2(n)}{\sigma_0^2+\sigma_1^2}
>\ln \left( 1+ \frac{\sigma_1^2}{\sigma_0^2}\right) \\
& z_0^2(n) >\frac{\sigma_0^2 (\sigma_0^2+\sigma_1^2)}{\sigma_1^2}
\ln \left( 1 +\frac{\sigma_1^2}{\sigma_0^2} \right) \\
&z_0^2(n)<z_1^2(n).
\end{split}
\label{A8}
\end{equation*}
These three conditions are equivalent to
\begin{equation*}
z_1^2(n) > z_0^2(n)  \;\text{and}\;  z_0^2(n) > \frac{\sigma_0^2
(\sigma_0^2+\sigma_1^2)}{\sigma_1^2} \ln \left( 1 +
\frac{\sigma_1^2}{\sigma_0^2} \right).
\end{equation*}


\section{Multivariate gamma distribution} \label{MultivariateGamma}
Define $p$ independent random vectors of $\mathbb{R}^2$ denoted as $\vv_k(\ell) = [v_0(\ell-k), v_1(\ell-k)]^\top\sim\mathcal{N}(\bb{0}, \bb{\Sigma})$, $k = 0,\dots, p-1$, and the $2\times p$ matrix $\bb{V}(\ell) = [\vv_0(\ell), \vv_1(\ell), \cdots, \vv_{p-1}(\ell)]$. 
Then, the $2\times 2$ matrix $\bb{A} = \bb{V}(\ell)\bb{V}^\top\!(\ell)$ is known to be distributed according to a Wishart distribution $\mathcal{W}_2(p,\bb{\Sigma})$ with $p$ degrees of freedom and covariance matrix $\bb{\Sigma}$~\cite[Theorem 3.2.4, p. 91]{muirhead2005}. 
Now, define the vector $\vd$ composed by the elements of the main diagonal of $\bb{A}$. Then, it was shown in Proposition 1.3.3 in~\cite[p. 32]{Chatelain_phd_2007} that $\vd$ is distributed according to a multivariate gamma distribution denoted $\mathcal{G}(q,P)$ with shape parameter $q=p/2$ and scale parameter $P = \{p_1,p_2,p_{12}\}$, with 
\begin{equation}
\begin{split}
&p_1 = 2\bb{\Sigma}(1,1)\\
&p_2 = 2\bb{\Sigma}(2,2)\\
&p_{12} = 4\left[\bb{\Sigma}(1,1)\bb{\Sigma}(2,2) -\bb{\Sigma}(1,2)\bb{\Sigma}(2,1)\right]
\end{split}
\label{eq:P}
\end{equation}
where $\bb{\Sigma}(1,1)$, $\bb{\Sigma}(1,2) = \bb{\Sigma}(2,1)$ and $\bb{\Sigma}(2,2)$ are the elements of the covariance matrix $\bb{\Sigma}$.

Now, making $\vv_k(n) = \vz(n-k) = [z_0(n-k), z_1(n-k)]^\top\sim\mathcal{N}(\bb{0}, \bb{\Sigma}_{ip})$, $k = 0,\dots,p-1$, for each hypothesis $\mathcal{H}_i$, and assuming the independence of $\vz(n-i)$ and $\vz(n-j)$ for $i \ne j$\footnote{This is a simplifying assumption employed for mathematical tractability. Simulation results will show that this assumption has little impact on the classifier performance.}, the above results show that $\vd(n) = [\|\vz_0(n)\|^2,\|\vz_1(n)\|^2]^\top$ is distributed according to a multivariate gamma distribution with shape parameter $q=p/2$ and scale parameter $P = \{p_1,p_2,p_{12}\}$ evaluated from \eqref{eq:P} with $\bb{\Sigma} = \bb{\Sigma}_{ip}$.

\section{Probability of error integrals}\label{app:PronIntegrals}
To lighten the notation, consider the vector $\vd(n) = [t_0,t_1]^\top$. Also consider $f_j$ to be the bivariate Gamma density associated to $\cH{j}$.
Thus, the probability of error $P_{ij}=P(\mathcal{H}_i|\mathcal{H}_j)$, can be computed as:

$P(\mathcal{H}_0|\mathcal{H}_j)$: $\mathcal{H}_0$ is accepted if $ t_1 < t_0$ and $t_1 < T$
  \begin{equation}
  \begin{split}
  P_{0j} &=\int_{0}^{T} \int_{t_1}^{+\infty} f_j(t_0,t_1) dt_0 dt_1\\
  &= \int_{0}^{T} \int_{0}^{+\infty} f_j(t_0+t_1,t_1) dt_0 dt_1
  \end{split}
  \label{eq:p0i}
  \end{equation}
  where the variable change $t_0' = t_0 + t_1$ was made.

$P(\mathcal{H}_1|\mathcal{H}_j)$: $\mathcal{H}_1$ is accepted if $ t_1 > t_0 \; \textrm{and} \; t_0 < T$,
  \begin{equation}
  \begin{split}
  P_{1j} &=\int_{0}^{T} \int_{t_0}^{+\infty} f_j(t_0,t_1) dt_1 dt_0\\
  &= \int_{0}^{+\infty} \int_{0}^{T} f_j(t_0,t_1+t_0) dt_0 dt_1
  \end{split}
  \end{equation}
  where the variable change $t_1' = t_1 - t_0$ was made.

$P(\mathcal{H}_2|\mathcal{H}_j)$: $\mathcal{H}_2$ is accepted if $t_1 < t_0 \; \textrm{and} \; t_1 > T,$
  \begin{equation}
  \begin{split}
  P_{2j} &=\int_{T}^{+\infty} \int_{t_1}^{+\infty} f_j(t_0,t_1) dt_0 dt_1\\
  &= \int_{T}^{+\infty} \int_{0}^{+\infty} f_j(t_0 + t_1,t_1) dt_0 dt_1
  \end{split}
  \end{equation}
  where the variable change $t_0' = t_0 - t_1$ was made.

$P(\mathcal{H}_3|\mathcal{H}_j)$: $\mathcal{H}_3$ is accepted if $ t_1 > t_0 \; \textrm{and} \; t_0 > T$
  \begin{equation}
  \begin{split}
  P_{3j} &=\int_{T}^{+\infty} \int_{t_0}^{+\infty} f_j(t_0,t_1) dt_1 dt_0\\
  &= \int_{0}^{+\infty} \int_{T}^{+\infty} f_j(t_0 ,t_1 +t_0) dt_0 dt_1
  \end{split}
  \label{eq:p3i}
  \end{equation}
  where the variable change $t_1' = t_1 - t_0$ was made.

\bibliographystyle{IEEEtran}
\bibliography{IEEEabrv,biblio}

\begin{biography}{Tales Imbiriba (S'14, M'17)}   
received the B.E.E. and M.Sc. degrees from the Federal University of Par\'a (UFPA), Bel\'em, Brazil, in 2006 and 2008, and his Doctorade degree from the Federal University of Santa Catarina (UFSC), Florian\'opolis, Brazil, in 2016. He is currently serving as a postdoctoral researcher at the Digital Signal Processing Laboratory (LPDS) at UFSC. 
His research interests include audio and image processing, pattern recognition, kernel methods, and adaptive filtering.
\end{biography}

\begin{biography}{Jos\'e Carlos M. Bermudez (S'78,M'85,SM'02)}
received the B.E.E. degree from Federal University of Rio de Janeiro
(UFRJ), Rio de Janeiro, Brazil, the M.Sc. degree in electrical
engineering from COPPE/UFRJ, and the Ph.D. degree in electrical
engineering from Concordia University, Montreal, Canada, in $1978$,
$1981$, $1985$, respectively.

He joined the Department of Electrical Engineering, Federal
University of Santa Catarina (UFSC), Florian\'opolis, Brazil, in
$1985$. He is currently a Professor of electrical engineering. In
the winter of $1992$, he was a Visiting Researcher with the
Department of Electrical Engineering, Concordia University. In
$1994$, he was a Visiting Researcher with the Department of
Electrical Engineering and Computer Science, University of
California, Irvine (UCI). His research interests have involved
analog signal processing using continuous-time and sampled-data
systems. His recent research interests are in digital signal
processing, including linear and nonlinear adaptive filtering,
active noise and vibration control, echo cancellation, image
processing, and speech processing.

Prof. Bermudez served as an Associate Editor for the IEEE
TRANSACTIONS ON SIGNAL PROCESSING in the area of adaptive filtering
from $1994$ to $1996$ and from $1999$ to $2001$, and as the Signal
Processing Associate Editor for the JOURNAL OF THE BRAZILIAN
TELECOMMUNICATIONS SOCIETY ($2005-2006$). He was a member of the
Signal Processing Theory and Methods Technical Committee of the IEEE
Signal Processing Society from $1998$ to $2004$. He is currently an
Associate Editor for the EURASIP JOURNAL ON ADVANCES IN SIGNAL
PROCESSING.
\end{biography}

\begin{biography}{Jean-Yves Tourneret}
(SM'08) received the ingénieur degree in electrical engineering from
\'Ecole Nationale Supérieure d'\'Electronique, d'\'Electrotechnique,
d'Informatique et d'Hydraulique of Toulouse (ENSEEIHT) in $1989$ and
the Ph.D. degree from the National Polytechnic Institute from
Toulouse in $1992$.

He is currently a professor in the university of Toulouse
(ENSEEIHT), France and a member of the IRIT laboratory (UMR $5505$
of the CNRS). His research activities are centered around
statistical signal processing with a particular interest to Markov
Chain Monte Carlo methods. He was the program chair of the European
conference on signal processing (EUSIPCO), which was held in
Toulouse (France) in $2002$. He was also member of the organizing
committee for the international conference ICASSP'06 which was held
in Toulouse (France) in $2006$. He has been a member of different
technical committees including the Signal Processing Theory and
Methods (SPTM) committee of the IEEE Signal Processing Society
($2001-2007$, $2010-present$). He is currently serving as an
associate editor for the IEEE TRANSACTIONS ON SIGNAL PROCESSING.
\end{biography}

\begin{biography}{Neil J. Bershad} received the B.E.E. degree from
Rensselaer Polytechnic Institute, Troy, NY, in $1958$, the M.S.
degree in electrical engineering from the University of Southern
California, Los Angeles, CA in $1960$, and the Ph.D.  degree in
electrical engineering from Rensselaer Polytechnic Institute in
$1962$. He joined the Faculty of the Henry Samueli School of
Engineering, University of California, Irvine in $1966$ and is now
an Emeritus Professor of Electrical Engineering and Computer
Science. His research interests have involved stochastic systems
modeling and analysis. His recent interests are in the area of
stochastic analysis of adaptive filters. He has published a
significant number of papers on the analysis of the stochastic
behavior of various configurations of the LMS adaptive filter. His
present research interests include the statistical learning behavior
of adaptive filter structures for echo cancellation, active acoustic
noise cancellation and variable gain (mu) adaptive algorithms. Dr.
Bershad has served as an Associate Editor of the IEEE TRANSACTIONS
ON COMMUNICATIONS in the area of phase-locked loops and
synchronization. More recently, he was an Associate Editor of the
IEEE TRANSACTIONS ON ACOUSTICS, SPEECH AND SIGNAL PROCESSING in the
area of adaptive filtering.
\end{biography}

\end{document}